\documentclass{pasj00}
\begin{document}
\SetRunningHead{Imanishi et al.}{Infrared L-band spectroscopy of QSOs}
%\Received{}%{yyyy/mm/dd}
%\Accepted{}%{yyyy/mm/dd}
%\Published{}%{yyyy/mm/dd}

\title{Infrared 3--4 $\mu$m Spectroscopy of Nearby PG QSOs and
AGN--Nuclear Starburst Connections in High-luminosity AGN Populations}

%%% begin:list of authors
% Do NOT capitalize all letters in "textsc".
\author{Masatoshi \textsc{Imanishi} %
\altaffilmark{1,2}}
\affil{Subaru Telescope, 650 North A'ohoku Place, Hilo, Hawaii, 96720,
U.S.A.}
\email{masa.imanishi@nao.ac.jp}

\author{Kohei \textsc{Ichikawa}, Tomoe \textsc{Takeuchi}}
\affil{Department of Astronomy, Kyoto University, Kyoto 606-8502, Japan}

\author{Nozomu \textsc{Kawakatu}}
\affil{Department of Physics, University of Tsukuba, Tennodai, Tsukuba
305-8577, Japan}

\author{Nagisa \textsc{Oi} \altaffilmark{2,3}, and Keisuke \textsc{Imase}
\altaffilmark{2,3}}
\affil{Department of Astronomy, School of Science, Graduate
University for Advanced Studies (SOKENDAI), Mitaka, Tokyo 181-8588, Japan}

\altaffiltext{1}{Department of Astronomy, School of Science, Graduate
University for Advanced Studies (SOKENDAI), Mitaka, Tokyo 181-8588, Japan}

\altaffiltext{2}{National Astronomical Observatory, 2-21-1, Osawa,
Mitaka, Tokyo 181-8588, Japan}

\altaffiltext{3}{Subaru Telescope, 650 North A'ohoku Place, Hilo,
Hawaii, 96720, U.S.A.}

%%% end:list of authors

%%% Please use the following style in case that sorting by 
%%% affilation is impossible. 
%
% \author{%
%   D-Firstname \textsc{D-Familyname}\altaffilmark{1}
%   E-Firstname \textsc{E-Familyname}\altaffilmark{1,2}
%   and
%   F-Firstname \textsc{F-Familyname}\altaffilmark{2}}
% \altaffiltext{1}{Address of Institute}
% \email{ddddd@xxx.xxx.xx.xx}
% \email{eeeee@xxx.xxx.xx.xx}
% \altaffiltext{2}{Address of Institute}

%% `\KeyWords{}' always has to be placed before `\maketitle'.
%Do NOT move this preamble from here! 

\maketitle

\begin{abstract}

We present the results of infrared $L$-band (3--4 $\mu$m) slit
spectroscopy of 30 PG QSOs at z $<$ 0.17, the representative sample of 
local high-luminosity, optically selected AGNs. 
The 3.3 $\mu$m polycyclic aromatic hydrocarbon (PAH) emission feature is
used to probe  
nuclear ($<$a few kpc) starburst activity and to investigate the
connections between AGNs and nuclear starbursts in PG QSOs. 
The 3.3 $\mu$m PAH emission is detected in the individual spectra of 5/30
of the observed PG QSOs. 
We construct a composite spectrum of PAH-undetected PG
QSOs and discern the presence of the 3.3 $\mu$m PAH emission therein.
We estimate the nuclear-starburst and AGN luminosities from the observed
3.3 $\mu$m PAH emission and 3.35 $\mu$m continuum luminosities,
respectively, and find that the nuclear-starburst-to-AGN luminosity ratios
in PG QSOs are similar to those of previously studied AGN populations with
lower luminosities, suggesting that AGN--nuclear starburst connections
are valid over the wide luminosity range of AGNs in the local universe.
The observed nuclear-starburst-to-AGN luminosity ratios in PG QSOs with 
available supermassive black hole masses are comparable to a theoretical
prediction based on the assumption that the growth of a supermassive
black hole is controlled by starburst-induced turbulence.

\end{abstract}

\section{Introduction}

The apparent ubiquity of supermassive black holes (SMBHs) in the
spheroidal components of present-day galaxies and the
correlation between the masses of SMBHs and spheroidal stars 
\citep{mag98,fer00} indicate that active galactic nucleus (AGN; 
SMBH-driven activity) and starburst activity are closely coupled.   
Theories also predict that starburst activity in the nuclear region of
a galaxy can have a strong effect on a central AGN by regulating the
mass accretion rate of a central SMBH
\citep{nor88,von93,ume97,ume98,ohs99,wad02,kaw03,vol03,kaw08,col08,wad09}.
Hence, it is important to observationally ascertain whether, in what 
types of galaxies, and on which physical scale AGN and starburst
activity are related. 

To disentangle AGN and starburst activity in galaxies, it is essential
to use an indicator that can trace {\it only} starburst or AGN activity. 
An effective tool for detecting starburst activity is the
polycyclic aromatic hydrocarbon (PAH) emission features found in the 
infrared 3--20 $\mu$m wavelength range \citep{tie08}.
In a starburst galaxy, PAHs in the photo-dissociation regions are excited
without destruction by far-UV photons from stars, and so a starburst
galaxy usually exhibits strong PAH emission features
\citep{moo86,roc91,gen98,imd00}.  
However, in the near vicinity of an AGN, PAHs are destroyed due to
strong X-ray radiation emanating from the AGN \citep{voi92}.
In the obscuring material surrounding a central AGN, PAH molecules at some
distance from the AGN could be adequately shielded from such X-ray emission
by gas and dust and could survive. 
However, no PAH-exciting UV photons from the AGN are available there,
as the extinction rate is greater in the UV range than in the X-ray range, 
and thus, PAH emission is absent from a pure AGN.  
If PAH emission is detected in a galaxy hosting an AGN, PAH-exciting 
UV photons from local energy sources (i.e., starbursts) are required. 
Thus, PAH emission is a very useful tool for detecting only starburst 
activity and investigating its properties by removing the contamination
from an AGN, even in an AGN/starburst composite galaxy.   
More importantly, since dust extinction is small at $\lambda >$ 3 $\mu$m 
($<$0.05 $\times$ A$_{\rm V}$; Nishiyama et al. 2008; 2009), the
absolute magnitude of modestly-obscured (A$_{\rm V}$ $<$ 20 mag) 
starburst activity is reasonably quantifiable from the 
{\it observed} (extinction-{\it uncorrected}) PAH emission
luminosity \citep{mou90,ima02,soi02,pee04,smi07}.  
For these reasons, infrared 3--20 $\mu$m spectroscopy can be very
effective for directly ascertaining AGN--starburst connections
in galaxies in a quantitatively reliable manner. 

Using the PAH emission features detected in the space-based Spitzer IRS
observations \citep{hou04}, AGN--starburst connections in optically
identified AGNs 
have been widely investigated \citep{sch06,shi07,wat08,wat09,wu09,vei09,bau10}. 
However, the large aperture of the Spitzer IRS ($>$3$\farcs$6) probes not
only nuclear ($<$a few kpc) starbursts, but also spatially extended 
($>$several kpc) star-forming regions in host galaxies. 
Although spatially extended star formation may be more quiescent than
nuclear starbursts, it can dominate the total PAH emission flux in
large-aperture 
spectroscopy because of its large surface area. 
With regard to possible connections between AGNs and starbursts,
however, it is likely that {\it nuclear} starbursts have a stronger influence
on an AGN due to the proximity to a central SMBH. 
The greater importance of {\it nuclear} starbursts (rather than
spatially extended star-formation) to central AGNs was actually
argued from observations \citep{wat08}, suggesting that it is important
to observationally investigate the relationship between {\it nuclear}
starbursts and AGNs. 

Ground-based, narrow-slit ($\sim$1'') spectroscopy is better suited
to detecting exclusively {\it nuclear} ($<$a few kpc) starbursts by
reducing the 
contamination from spatially extended ($>$several kpc), quiescent
star-forming activity in host galaxies.  
The 3.3 $\mu$m PAH emission feature is particularly useful for probing 
exclusively nuclear starbursts via ground-based slit spectroscopy, 
as it is in an appropriate wavelength range for terrestrial atmospheric
transmission (i.e., $L$-band; 2.8--4.2 $\mu$m).
Although a strong 11.3 $\mu$m PAH emission feature is also
observable from the ground, the Earth's atmospheric background level 
in the $N$-band (8--13 $\mu$m) is much higher than that in the $L$-band,
making it more difficult to obtain spectra with sufficient S/N ratios around
the 11.3 $\mu$m PAH emission.  

Using the 3.3 $\mu$m PAH emission features detected in ground-based,
narrow-slit spectra, the relationship between AGNs and {\it nuclear}
starbursts was observationally investigated in the nearby Seyfert galaxies
(a moderate-luminosity AGN population), and a positive luminosity
correlation was found \citep{ima02,ima03,rod03,iw04}. 
\citet{oi10} has extended these results to low-luminosity AGNs and 
has confirmed that a similar correlation is valid in this case. 
It would be interesting to know if the AGN--nuclear-starburst
connection holds in a high-luminosity AGN population, as
theoretical predictions about AGN--nuclear-starburst connections are
varied, and are sometimes contradictory. 
For example, \citet{kaw08} predicted that the nuclear-starburst-to-AGN 
luminosity ratio will increase with increasing AGN luminosity, whereas
\citet{bal08} has argued the contrary. 
Observations of high-luminosity AGNs are needed to resolve this issue.

In this paper, we present the results of ground-based, narrow-slit
spectroscopy of PG QSOs \citep{sch83}, a higher-luminosity AGN population
than the Seyfert galaxies in the local universe. 
Throughout this paper, $H_{0}$ $=$ 75 km s$^{-1}$ Mpc$^{-1}$,
$\Omega_{\rm M}$ = 0.3, and $\Omega_{\rm \Lambda}$ = 0.7 are
adopted to be consistent with our previous publications. 

\section{Targets}

Target objects were selected from the PG QSO sample compiled by
\citet{sch83}, which is the representative high-luminosity,
optically selected, type 1 unobscured AGN population in the local
universe. 
Sources with M$_{\rm B}$ $<$ $-$23 mag, conventionally classified as 
QSOs \citep{sch83}, are the objects of particular interest for investigating 
AGN--nuclear-starburst connections in the high-luminosity AGN range.
For the redshifted 3.3 $\mu$m PAH emission to be observable at a 
wavelength appropriate for terrestrial atmospheric transmission (i.e.,
$L$-band; 2.8--4.2 $\mu$m), we limit the redshifts of the
target objects to less than 0.17. 
We divide the objects according to whether M$_{\rm B}$ $<$ $-$24.1 mag
or $-$24.1 $\leq$ M$_{\rm B}$ $<$ $-$23 mag (in the cosmology of Schmidt
\& Green, 1983), and refer to these two classes as high- and
low-luminosity PG QSOs, respectively.   
The number of low-luminosity PG QSOs is larger than the number of
high-luminosity PG QSOs.

For the high-luminosity PG QSOs, we set a declination limit of
$-$15$^{\circ}$ $\sim$ $+$65$^{\circ}$ to facilitate observation with 
small airmass at Mauna Kea, Hawaii (our observation site). 
This selection resulted in ten high-luminosity PG QSOs at $z <$ 0.17
(Table 1), all of which were observed. 

Because the number of low-luminosity PG QSOs is larger, a
more stringent declination limit of $-$15$^{\circ}$ $\sim$
$+$55$^{\circ}$ was imposed. 
Twenty low-luminosity PG QSOs remained (Table 1), and all were 
observed. 
 
Our sample is statistically complete, and hence is useful for investigating 
general properties without any obvious selection bias. 
The redshift range is 0.061--0.167, where 1'' corresponds to 
the physical scale of 1.1--2.7 kpc.

\section{Observations and Data Reduction}

We used the IRCS infrared spectrograph \citep{kob00} at the 
Nasmyth focus of the Subaru telescope \citep{iye04} to obtain the new
infrared $L$-band spectra of the PG QSOs, with the exception of Mrk 1014,
3C 273, and I Zw 1, for which the IRTF telescope was used. 
Table 1 lists the observational details.

During the Subaru IRCS observations, the sky was clear.
The seeing size in the $K$-band ($\lambda$ = 2.2 $\mu$m), obtained from 
images taken
prior to the $L$-band spectroscopy, was 0$\farcs$4--0$\farcs$8 in FWHM. 
The 0$\farcs$9-wide slit and $L$-grism were used with a 52-mas-pixel scale. 
The achievable spectral resolution was R $\sim$ 140 at $\lambda \sim$ 
3.5 $\mu$m. 
The precipitable water was very low, 1--2 mm for the 2006 July and 2010
April observational runs, and $<$1 mm for the 2007 January run. 
We adopted a standard telescope nodding technique (ABBA pattern) with a
throw of 7'' along the slit to subtract background emission.
We used the optical guider of the Subaru telescope to monitor the
telescope tracking. 
The exposure time was 1.0--1.8 sec, and 30--60 coadds were employed at each
nod position. 

The IRCS was moved from the Cassegrain to the Nasmyth focus of the Subaru
telescope in 2005. 
Afterward, the background level was high in the $L$-band
because of the high emissivity of the Nasmyth image rotator.  
In 2008, the image rotator was recoated, and the resulting low
background level enabled a higher achievable sensitivity in the 
2010 April data than in the 2006--2007 data.
Hence, for objects of the same apparent magnitude observed with the
same exposure time, the data quality of the 2010 spectra
is generally superior to that of the 2006--2007 spectra. 

For all observational runs, F- or G-type main sequence stars
(Table 1) were observed as standard stars, with a mean airmass
difference of $<$0.1 compared with the individual PG QSOs, to 
correct for terrestrial atmospheric transmission and provide flux
calibration. The $L$-band magnitudes of the standard stars were
estimated from their $V$-band ($\lambda =$ 0.6 $\mu$m) magnitudes,
adopting the $V-L$ colors appropriate for the stellar types of the
individual standard stars \citep{tok00}.

Standard data analysis procedures were employed using IRAF   
\footnote{IRAF is distributed by the National Optical Astronomy
Observatories, operated by the Association of Universities
for Research in Astronomy, Inc. (AURA), under cooperative agreement
with the National Science Foundation.}.
%--------------
Initially, frames taken with an A (or B) beam were subtracted from
frames subsequently taken with a B (or A) beam, and the resulting
subtracted frames were added and divided by a spectroscopic flat
image. Bad pixels and pixels impacted by cosmic rays were then replaced
with the interpolated values of the surrounding pixels. Finally, the
spectra of the PG QSOs and the standard stars were extracted. 
The wavelength calibration was based on wavelength-dependent
terrestrial atmospheric transmission. 
The spectra of the PG QSOs were divided by the observed spectra of the
standard stars and were multiplied by the spectra of blackbodies with 
temperatures appropriate to the individual standard stars (Table 1).

The flux calibration was based on the signals from the PG QSOs and
the standard stars detected inside our slit spectra. 
Seeing sizes in the $K$-band (and $L$-band) were always smaller
than the employed slit widths, and the tracking performance
of the Subaru telescope was already established. 
We thus expected that the possible slit loss for the compact emission 
would be insignificant.  
To reduce the scatter of the spectra, appropriate binning of spectral
elements was performed, particularly for faint sources.

Mrk 1014 and I Zw 1 were observed with the IRTF SpeX \citep{ray03},
and 3C 273 was observed with the IRTF NSFCAM \citep{shu94}.
The details of the Mrk 1014 and 3C 273 observations are found in
\citet{idm06}.
For I Zw 1, the observational details and data reduction were carried
out in the 
same way as with Seyfert galaxies observed during the same period (2003 
September) \citep{iw04}.
In summary, the 1$\farcs$6-wide slit and the 1.9--4.2 $\mu$m
cross-dispersed mode of the SpeX were employed.
The effective spectral resolution was R $\sim$ 500.

\section{Results}

Figures 1 and 2 display the respective infrared 3.4--3.9 $\mu$m spectra 
of the high- and low-luminosity PG QSOs.
The spectra of Mrk 1014 and 3C 273 were originally published in \citet{idm06}, 
and are reproduced here.  
In all sources, the redshifted 3.3 $\mu$m PAH emission is within the 
3.4--3.9 $\mu$m wavelength range in the observed frame 
($\lambda_{\rm obs}$). 
Only the spectra in the range $\lambda_{\rm obs}$ = 3.4--3.9 $\mu$m are shown,
as
(1) the data for $\lambda_{\rm obs}$ $<$ 3.4 $\mu$m generally exhibit large
scatter due to the strong wavelength dependency of terrestrial atmospheric
transmission, and (2) the data for $\lambda_{\rm obs}$ $>$ 3.9 $\mu$m are 
noisier owing to the increased terrestrial atmospheric background emission.
This treatment is necessary to better visualize the possible signatures
of weak 3.3 $\mu$m PAH emission features in PG QSOs. 

Photometry at 3.7 $\mu$m with a 5''--5$\farcs$5 aperture can be found in the
literature for many of the observed PG QSOs \citep{neu87,san89}, and our
flux estimates roughly agree with the photometric flux measurements in
most cases, suggesting that (1) the infrared 3--4 $\mu$m emission is
dominated by a spatially-compact component, and (2) the slit loss in our
slit spectra is generally insignificant.  

Unlike the Seyfert galaxies, which contain both type 1 unobscured and type 2
obscured AGNs \citep{huc92,spi89}, the high-luminosity AGN sample is
well established only for type 1 unobscured AGNs \citep{sch83}.
For PG QSOs (=type 1 unobscured AGNs), the unattenuated 
featureless continuum emission originating from an AGN reduces the 
equivalent width of the 3.3 $\mu$m PAH emission originating from a starburst.
Consequently, even for the PG QSOs with 3.3 $\mu$m PAH emission
signatures, the PAH flux excess above the continuum level is small, making 
it difficult to trace the 3.3 $\mu$m PAH emission profile in detail.
We thus assume a typical profile for the 3.3 $\mu$m PAH emission 
feature \citep{tok91}.
Table 2 summarizes the flux (f$_{\rm 3.3PAH}$), luminosity (L$_{\rm
3.3PAH}$), and rest-frame equivalent width 
(EW$_{\rm 3.3PAH}$) of the 3.3 $\mu$m PAH emission for the five
PAH-detected PG QSOs. 

For comparison with 3.3 $\mu$m PAH-derived starburst magnitudes, AGN
luminosities are estimated from the continuum emission 
at $\lambda_{\rm rest}$ = 3.35 $\mu$m.  
Given that the rest-frame equivalent width of the 3.3 $\mu$m PAH emission
feature (Table 2) is more than an order of magnitude lower than that of a pure
starburst-dominated galaxy ($\sim$100 nm; Moorwood 1986; Imanishi \&
Dudley 2000), it is reasonable to assume that the observed 
$\lambda_{\rm rest}$ = 3.35 $\mu$m continuum emission of PG QSOs 
originates predominantly from AGNs. 
Table 3 lists the AGN continuum luminosities ($\lambda$L$_{\lambda}$),
and Figure 3(a) shows a histogram of the luminosities.
The AGN continuum luminosity traces hot dust emission heated by the AGN
at the inner part of the putative dusty torus \citep{ant85,bar87}.

Our spectra include important information on the continuum slopes of 
high-luminosity, type 1 unobscured AGNs. 
We estimate the continuum slope $\Gamma$ (F$_{\lambda}$ $\propto$ 
$\lambda^{\Gamma}$) from data points in the range $\lambda_{\rm rest}$ 
= 3.4--3.9 $\mu$m, as summarized in Table 3. 
The histogram of the $\Gamma$ values is shown in Figure 3(b).
A larger $\Gamma$ value means a redder continuum.
The $\Gamma$ values are in the range $-$1.2 $\sim$ $-$0.2, except for
PG 0052+251 and PG 1202+281. 
The median $\Gamma$ value is $-$0.8, which is larger (redder) 
than for starburst-dominated galaxies ($\Gamma$ = $-$2; F$_{\lambda}$ 
$\propto$ $\lambda^{\Gamma}$; Imanishi et al. 2010). 

\section{Discussion}

\subsection{AGN - nuclear starburst connections in PG QSOs}

The detection rate of the 3.3 $\mu$m PAH emission feature 
is only 17\% (= 5/30) for PG QSOs, suggesting that the relative
contribution from starburst emission is generally weak in the 3--4
$\mu$m range when compared with the continuum emission originating from
an AGN.   
This is what we expected, as signs of the 3.3 $\mu$m PAH
emission feature are not obvious in the composite spectrum of SDSS
QSOs (=high-luminosity, optically selected, type 1 unobscured AGNs)
\citep{gli06}. 
Given the high fraction of PAH-undetected sources, it is difficult 
to impose strong constraints on starburst properties based on
the upper limits of PAH strength in individual spectra. 
Instead, we construct a composite spectrum of PAH-undetected
PG QSOs to discuss the AGN--nuclear-starburst luminosity ratios in a
more quantitatively detailed manner.
Among the 25 PG QSOs with not clearly detectable 3.3 $\mu$m PAH emission in
the individual spectra, we choose the 23 observed with the Subaru IRCS 
(i.e., we exclude I Zw 1 and 3C 273) to eliminate systematic errors
caused by the inclusion of data obtained with different instruments. 
We normalize the flux level based on the signals in the range
$\lambda_{\rm rest}$ 
= 3.1--3.5 $\mu$m and construct a composite spectrum in which all objects have 
equal weight using the IRAF task {\it sarith}.

Figure 4 displays the composite spectrum, in which the signature of the
3.3 $\mu$m PAH emission is barely seen. 
By assuming the typical 3.3 $\mu$m PAH profile (type A of Tokunaga et
al., 1991), we estimate the 3.3 $\mu$m PAH emission strength.
The excess above the continuum is 2.1\%, and the peak wavelength is 
$\lambda_{\rm rest}$ $\sim$ 3.29 $\mu$m.
The rest-frame 3.3 $\mu$m PAH equivalent width is estimated to be
$\sim$1 [nm] (Table 2), assuming a median redshift of $\sim$0.1. 
The luminosity ratio of 3.3 $\mu$m PAH emission to AGN-originated 
3.35$\mu$m continuum ($\lambda$L$_{\lambda}$) is 3.5 $\times$ 10$^{-4}$.

We here regard that the detected PAH emission toward the nuclear
directions of PG QSOs is dominated by starburst activity in the true 
{\it nuclear regions}, rather than just a small fraction of
spatially-extended star-forming activity, which happens to occur at the
foreground of the nuclear directions, but is physically far away from the
AGN nuclei, for the following reason.
Our PG QSO sample is optically classified as type-1 unobscured AGNs, and
so AGN obscuration is very small in the optical. 
Stars are formed from molecular gas, where dust usually coexists. 
If a large amount of star-forming activity in an extended host galaxy's
disk is present in front of the nuclear AGN direction, such a galaxy 
would not be classified as a type-1 AGN in the optical. 
Hence, the inclination of extended star-forming disks in our PG QSO
sample must be biased to face-on \citep{wu07}, where the contamination
from star-forming activity at a large physical distance from the nuclei,
to the observed nuclear spectra, is minimal. 

\subsection{Comparison with a lower-luminosity AGN population}

Figure 5 compares the 3.3 $\mu$m PAH emission luminosity (ordinate) and the
$\lambda_{\rm rest}$ = 3.35 $\mu$m continuum luminosity in 
$\lambda$L$_{\lambda}$ (abscissa) for PG QSOs.
The ordinate and abscissa trace the nuclear-starburst and
AGN-heated hot dust luminosities, respectively. 
The five PAH-detected PG QSOs are superimposed on the data points by 
\citet{iw04} and \citet{oi10}.
The data point of the composite spectrum of the 23 PAH-undetected
PG QSOs is also superimposed, where the median AGN continuum
luminosity of 
$\lambda$L$_{\lambda}$(3.35 $\mu$m) $\sim$ 3 $\times$ 10$^{44}$ 
[ergs s$^{-1}$] for the 23 objects is adopted in the abscissa.  
Figure 5 shows that (1) the data points of the PG QSOs are distributed in
a higher AGN luminosity range compared with the Seyfert galaxies
previously studied by \citet{iw04} and \citet{oi10}, and (2) the
nuclear-starburst-to-AGN luminosity  
ratios of PG QSOs are similar to or possibly higher
than those of Seyfert galaxies (=lower-luminosity AGNs).
The hypothesis that the nuclear-starburst-to-AGN luminosity ratio decreases 
with increasing AGN luminosity \citep{bal08} is therefore not supported. 
If the apparently higher ratio in PG QSOs is real, the model developed by
\citet{kaw08} may be preferable, but more detailed discussions are
difficult with our currently available dataset. 
 
To compare Seyfert galaxies and PG QSOs, it should be
noted that the AGN-heated hot dust emission luminosities for Seyferts
are derived from $N$-band (10 $\mu$m) continuum luminosities, whereas
those for PG QSOs are obtained from $L$-band (3.35 $\mu$m) continuum
luminosities.  
The reasons for this are as follows:
(1) Unlike the case for Seyfert galaxies, for which $N$-band
photometric measurements with small ($<$1--2'') apertures are 
available for many sources (e.g., Gorjian et al. 2004), such
measurements are not available for the observed PG QSOs. 
(2) Only type 1 unobscured AGNs are included in the PG QSO sample, so
that dust extinction effects are not a serious concern at wavelengths
shorter than the $N$-band.  
Both $L$- and $N$-band continuum luminosities trace the AGN-heated hot dust
emission.  
However, the shorter $L$-band wavelength preferentially probes the hotter dust 
closer to the central mass-accreting SMBH at the inner part of the
dusty torus compared with the longer $N$-band wavelength.
If the solid angle of the dusty torus, viewed from the central SMBH, does
not vary with the distance from the SMBH, then the $L$- and $N$-band
continuum luminosities should be similar, as the luminosity ratio
of the intrinsic energetic radiation of the AGN to the AGN-heated hot dust
luminosity depends only on the covering factor of the dusty torus.  
However, if the torus is flared \citep{wad02} or warped \citep{san89}, 
the AGN-originated $N$-band-to-$L$-band luminosity ratio could be greater
than unity. 
If this is the case, the true distribution of PG QSOs should move 
to the right relative to the current plots. 
Although this is a possible ambiguity, we do not expect it to be significant,
as both the $L$- and $N$-band continua probe relatively hot dust
($\sim$1000K and $\sim$300K, respectively) at the inner part of the
dusty torus, and thus, the effects of possible torus flaring/warping should
be smaller than with more widely separated wavelengths such
as $L$-band and $>$20 $\mu$m (which probes cooler dust).
In fact, we compared the 3.7 $\mu$m and 10.1 $\mu$m continuum luminosities 
($\lambda$L$_{\lambda}$) measured with a $\sim$5'' aperture for PG
QSOs with available measurements of this type \citep{neu87,san89}, but
could find no 
evidence that the 10.1 $\mu$m continuum luminosity is systematically larger
by a factor of $>$2 than the 3.7 $\mu$m continuum luminosity.

Finally, the bolometric luminosity of nuclear starbursts can be roughly
estimated from the observed 3.3 $\mu$m PAH emission luminosity using
the 3.3 $\mu$m PAH-to-infrared luminosity ratio of
$\sim$10$^{-3}$ for starbursts \citep{mou90,ima02} and assuming that
infrared luminosity dominates the bolometric luminosity in starbursts. 
To derive the AGN bolometric luminosity from the $L$-band continuum
luminosity ($\lambda$L$_{\lambda}$ or $\nu$L$_{\nu}$), we adopt a
correction factor of 5 \citep{ris10}. 
The dotted line in Figure 5 represents the nuclear-starburst-to-AGN 
bolometric luminosity ratio of 0.1. 
Based on the assumption that starburst-driven turbulence controls the
mass accretion of a central SMBH, \citet{kaw08} argued that the 
nuclear-starburst-to-AGN luminosity ratio tends to increase with increasing
final supermassive black hole mass (M$_{\rm SMBH}$), and the ratio can be
$>$0.1 if the mass is $>$10$^{8}$M$_{\odot}$. 
Table 3 (column 4) lists the M$_{\rm SMBH}$ values of the observed PG
QSOs, and the histogram is displayed in Figure 3(c). 
Four of the five PAH-detected PG QSOs (filled stars in Figure 5) 
with high nuclear-starburst-to-AGN 
luminosity ratios have M$_{\rm SMBH}$ $>$ 10$^{7.8}$M$_{\odot}$. 
The composite spectrum of the 23 PAH-undetected PG QSOs with median 
M$_{\rm SMBH}$ $\sim$ 10$^{8}$M$_{\odot}$ has marginal PAH detection,
and the implied nuclear-starburst-to-AGN luminosity ratio is $\sim$0.1.
These overall observational behaviors are reproduced by the model of
\citet{kaw08}.  

\section{Summary}

We reported the results of ground-based infrared 3--4 $\mu$m slit
spectroscopy of 30 PG QSOs. 
The 3.3 $\mu$m PAH emission feature was used to probe the magnitude of
nuclear starburst activity. 
The following are our primary conclusions.

\begin{enumerate}

\item We detected 3.3 $\mu$m PAH emission features (an effective
means of probing starburst activity) in the individual spectra of five
sources.  
For PAH-undetected PG QSOs, our composite spectrum revealed the
signature of the 3.3 $\mu$m PAH emission feature. 
The 3.3 $\mu$m PAH-derived nuclear-starburst-to-AGN luminosity ratios
were found to be comparable to, or possibly slightly higher than, those
of previously investigated lower-luminosity AGNs, suggesting that 
AGN--nuclear starburst connections hold in the wide luminosity range of
AGNs.  

\item The continuum slopes $\Gamma$ (F$_{\lambda}$ $\propto$
$\lambda^{\Gamma}$) of the PG QSOs were mostly $-$1.2 $\sim$ $-$0.2, 
with a median value of $-$0.8. 
This is larger (redder) than in starburst galaxies (median = $-$2).

\item The implied nuclear-starburst-to-AGN luminosity ratios 
for PG QSOs and their trend as a function of AGN luminosity 
could be explained by the theoretical model of \citet{kaw08}, which
assumes that SMBH growth is controlled by starburst-induced turbulence.

\end{enumerate}

We are grateful to Subaru and IRTF staff members for their support during our
observational runs. 
We thank the anonymous referee for his/her useful comments.
This work is based data collected at Subaru Telescope, which is operated
by the National Astronomical Observatory of Japan.
MI is supported by a Grant-in-Aid for Scientific Research (no. 22012006).
This research made use of the SIMBAD database operated by CDS,
Strasbourg, France, and the NASA/IPAC Extragalactic Database
(NED), which is operated by the Jet Propulsion Laboratory, California
Institute of Technology, under contract with the National Aeronautics
and Space Administration.

\clearpage

%%%%%%%%%% Table 1 %%%%%%%%%
\begin{table}[h]
\scriptsize
\caption{Target and Observation Log \label{tbl-1}}
\begin{center}
\begin{tabular}{lcllcrlccc}
\hline
\hline
Object & Redshift & Date & Telescope & Exp & P.A. & 
\multicolumn{4}{c}{Standard Stars} \\
 &  & (UT) & Instrument & (Min) & ($^{\circ}$) & Name &  
$L$-mag & Type & $T_{\rm eff}$ (K) \\
(1) & (2) & (3) & (4) & (5) & (6) & (7) & (8) & (9) & (10) \\ 
\hline
PG 0026+129 & 0.142 & 2007 January 14 & Subaru IRCS & 48 & 90 & HR 145 &
5.1 & F7V & 6240 \\ 
PG 0052+251 & 0.155 & 2007 January 15 & Subaru IRCS & 32 & 90 & HR 217 &
5.2 & F8V & 6000 \\ 
PG 0157+001 (Mrk 1014) & 0.164 & 2003 September 8 & IRTF SpeX & 160 & 0
& HR 650  & 4.1 & F8V & 6000  \\
PG 1048+342 & 0.167 & 2007 January 14 & Subaru IRCS & 56 & 90 & HR 4027
& 5.0 & G0V & 5930 \\
PG 1202+281 & 0.165 & 2007 January 14 & Subaru IRCS & 24 & 90 & HR 4496
& 3.5 & G8V & 5400 \\
PG 1226+023 (3C 273) & 0.158 & 2004 Apr 5 & IRTF NSFCAM & 28 & 0 & HR
4708 & 5.0 & F8V & 6000 \\ 
PG 1307+085 & 0.155 & 2007 January 15 & Subaru IRCS & 32 & 90 & HR 4708
& 5.0 & F8V & 6000 \\
PG 1352+183 & 0.158 & 2006 July 19 & Subaru IRCS & 32 & 90 & HR 4926 &
5.0 & F6V & 6400 \\ 
PG 1402+261 & 0.164 & 2006 July 19 & Subaru IRCS & 32 & 90 & HR 4926 &
5.0 & F6V & 6400 \\ 
PG 1613+658 & 0.129 & 2006 July 17 & Subaru IRCS & 24 & 90 & HR 6360 &
4.5 & G5V & 5700 \\ 
\hline
PG 0050+124 (I Zw 1) & 0.061 & 2003 September 8 & IRTF SpeX & 40 & 0 &
HR 145 & 5.1 & F7V & 6240 \\
PG 0844+349 & 0.064 & 2007 January 14 & Subaru IRCS & 40 & 90 & HR 3451
& 5.0 & F7V & 6240 \\
PG 1001+054 & 0.161 & 2007 January 15 & Subaru IRCS & 40 & 90 & HR 4079
& 5.3 & F6V & 6400 \\
PG 1114+445 & 0.144 & 2010 April 5 & Subaru IRCS & 64.8 & 90 & HR 4285 &
4.6 & F9V & 6000 \\ 
PG 1115+407 & 0.154 & 2010 April 4 & Subaru IRCS & 64.8 & 90 & HR 4285 &
4.6 & F9V & 6000 \\ 
PG 1211+143 & 0.085 & 2007 January 15 & Subaru IRCS & 32 & 90 & HR 4708
& 5.0 & F8V & 6000 \\
PG 1229+204 & 0.064 & 2010 April 4 & Subaru IRCS & 28.8 & 90 & HR 4864 &
4.5 & G7V & 5500 \\ 
PG 1404+226 & 0.098 & 2010 April 5 & Subaru IRCS & 57.6 & 90 & HR 5243 &
4.9 & F6V & 6400 \\ 
PG 1411+442 & 0.089 & 2007 January 15 & Subaru IRCS & 12 & 90 & HR 5423
& 4.7 & G5V & 5700 \\
PG 1415+451 & 0.114 & 2010 April 5 & Subaru IRCS & 57.6 & 90 & HR 5423 &
4.7 & G5V & 5700 \\ 
PG 1416$-$129 & 0.129 & 2006 July 20  & Subaru IRCS & 32 & 90 & HR 5322
& 4.9 & F9V & 6000 \\
PG 1426+015 & 0.086 & 2010 April 4 & Subaru IRCS & 57.6 & 90 & HR 5307 &
5.1 & F7V & 6240 \\ 
PG 1435$-$067 & 0.129 & 2006 July 20 & Subaru IRCS & 32 & 90 & HR 5322 &
4.9 & F9V & 6000 \\ 
PG 1440+356 & 0.077 & 2010 April 5 & Subaru IRCS & 82.8 & 90 & HR 5423 &
4.7 & G5V & 5700  \\  
PG 1519+226 & 0.137 & 2006 July 18 & Subaru IRCS & 32 & 90 & HR 5728 &
4.5 & G3V & 5800 \\ 
PG 1552+085 & 0.119 & 2010 April 4 & Subaru IRCS & 57.6 & 90 & HR 5659 &
5.1 & G5V & 5700 \\ 
PG 1612+261 & 0.131 & 2006 July 20 & Subaru IRCS & 48 & 90 & HR 6064 &
5.2 & G1V & 5900 \\ 
PG 1617+175 & 0.114 & 2006 July 17 & Subaru IRCS & 24 & 90 & HR 6064 &
5.2 & G1V & 5900 \\ 
PG 2130+099 & 0.061 & 2006 July 18 & Subaru IRCS & 24 & 90 & HR 8077 &
4.6 & F8V & 6000 \\ 
PG 2214+139 & 0.067 & 2006 July 19 & Subaru IRCS & 32 & 90 & HR 8955 &
4.7 & G0V & 5930 \\ 
\hline  
\end{tabular}
\end{center}
\end{table}

Notes: 

Col. (1): Object name. High-luminosity PG QSOs are listed in the first
ten rows. 

Col. (2): Redshift. 

Col. (3): Observation date in UT.

Col. (4): Telescope and instrument. 

Col. (5): Net on-source integration time in min.

Col. (6): Position angle of the slit. The north-south direction is
defined as 0$^{\circ}$.

Col. (7): Standard star name.

Col. (8): Adopted $L$-band magnitude of standard star.

Col. (9): Stellar spectral type of standard star.

Col. (10): Effective temperature of standard star.

\clearpage

\normalsize

%---- Table 2 ----%
\begin{table}[h]
%\small
\scriptsize
\caption{Strength of the 3.3 $\mu$m PAH Emission for Detected
Sources \label{tab2}} 

\begin{center}
\begin{tabular}{lccc}
\hline
\hline
Object & f$_{3.3 \rm PAH}$ & L$_{3.3 \rm PAH}$ & rest EW$_{\rm 3.3PAH}$ \\ 
 & [$\times$10$^{-14}$ ergs s$^{-1}$ cm$^{-2}$] & 
[$\times$10$^{41}$ ergs s$^{-1}$] & [nm] \\ 
(1) & (2) & (3) & (4)  \\
\hline
Mrk 1014 & 3.0$\pm$0.2 & 21$\pm$1.2 & 7 \\
PG 1211+143 & 1.7$\pm$0.5 & 2.7$\pm$0.8 & 1 \\ 
PG 1411+442 & 1.8$\pm$0.3 & 3.1$\pm$0.6 & 2 \\
PG 1416$-$129 & 0.66$\pm$0.2 & 2.5$\pm$0.7 & 5 \\ 
PG 1440+356 & 2.5$\pm$0.4 & 3.2$\pm$0.5 & 2 \\ 
Composite & --- & --- & 1 \\
\hline  
\end{tabular}
\end{center}
\end{table}

Notes.

Col.(1): Object name.

Col.(2): Observed flux of 3.3 $\mu$m PAH emission in 
[10$^{-14}$ ergs s$^{-1}$ cm$^{-2}$].

Col.(3): Observed luminosity of 3.3 $\mu$m PAH emission 
in [10$^{41}$ ergs s$^{-1}$].

Col.(4): Rest-frame equivalent width of 3.3 $\mu$m PAH emission 
in [nm].

%%%%%%%%%% Table 3 %%%%%%%%%
\begin{table}[h]
\scriptsize
\caption{AGN Luminosity and Continuum Slope \label{tbl-3}}
\begin{center}
\begin{tabular}{lccc}
\hline
\hline
Object & L$_{\rm L}$(AGN) & Continuum & M$_{\rm SMBH}$ \\
 & [$\times$ 10$^{44}$ ergs s$^{-1}$] & Slope ($\Gamma$) & [M$_{\odot}$]  \\
(1) & (2) & (3) & (4) \\ 
\hline
PG 0026+129 & 3.9 & $-$0.9 & 7.85 \\
PG 0052+251 & 4.2 & $-$1.6 & 8.72 \\
PG 0157+001 (Mrk 1014) & 9.1 & $-$0.5 & 8.10 \\
PG 1048+342 & 1.6 & $-$0.3 & 8.25 \\
PG 1202+281 & 4.4 & $+$0.3 & 8.30 \\
PG 1226+023 (3C 273) & 63.6 & $-$0.9 & 9.00 \\
PG 1307+085 & 5.7 & $-$0.8 & 7.86 \\
PG 1352+183 & 3.2 & $-$0.3 & 8.27 \\
PG 1402+261 & 0.9 & $-$0.3 & 7.30 \\
PG 1613+658 & 2.3 & $-$0.3 & 8.99 \\
\hline
PG 0050+124 (I Zw 1) & 5.7 & $-$0.8 & 7.26  \\
PG 0844+349 & 1.9 & $-$1.0 & 7.69 \\
PG 1001+054 & 7.7 & $-$1.1 & 7.63 \\
PG 1114+445 & 6.5 & $-$0.3 & 8.41 \\
PG 1115+407 & 6.6 & $-$1.1 & --- \\
PG 1211+143 & 6.8 & $-$0.4 & 7.88 \\
PG 1229+204 & 0.9 & $-$0.7 & 7.93 \\
PG 1404+226 & 1.2 & $-$0.9 & --- \\
PG 1411+442 & 5.2 & $-$0.4 & 7.89 \\
PG 1415+451 & 3.2 & $-$0.9 & 7.81 \\
PG 1416$-$129 & 1.4 & $-$0.2 & 8.50 \\
PG 1426+015 & 4.8 & $-$1.0 & 8.75 \\
PG 1435$-$067 & 2.8 & $-$0.4 & 8.24 \\
PG 1440+356 & 5.1 & $-$0.8 & 7.30 \\
PG 1519+226 & 2.2 & $-$0.8 & 7.78 \\
PG 1552+085 & 2.6 & $-$1.1 & --- \\
PG 1612+261 & 1.2 & $-$1.1 & 7.91 \\
PG 1617+175 & 1.4 & $-$1.2 & 8.70 \\
PG 2130+099 & 3.2 & $-$0.8 & 7.68 \\
PG 2214+139 & 1.9 & $-$0.7 & 8.42 \\
\hline  
\end{tabular}
\end{center}
\end{table}

Notes: 

Col. (1): Object name.

Col. (2): AGN luminosity derived from continuum luminosity 
at $\lambda_{\rm rest}$ = 3.35 $\mu$m ($\lambda$L$_{\lambda}$) in 
10$^{44}$ [ergs s$^{-1}$]. 

Col. (3): Continuum slope $\Gamma$ (F$_{\lambda}$ $\propto$
$\lambda^{\Gamma}$).  

Col. (4): Mass of supermassive blackhole in units of solar mass, adopted
from \citet{kaw07} and \citet{wat09}, after conversion to our
cosmology. For Mrk 1014, we adopt the estimate of \citet{wat09}. 

\clearpage

%---  Figure 1 ---%
\begin{figure}
\includegraphics[angle=-90,scale=.185]{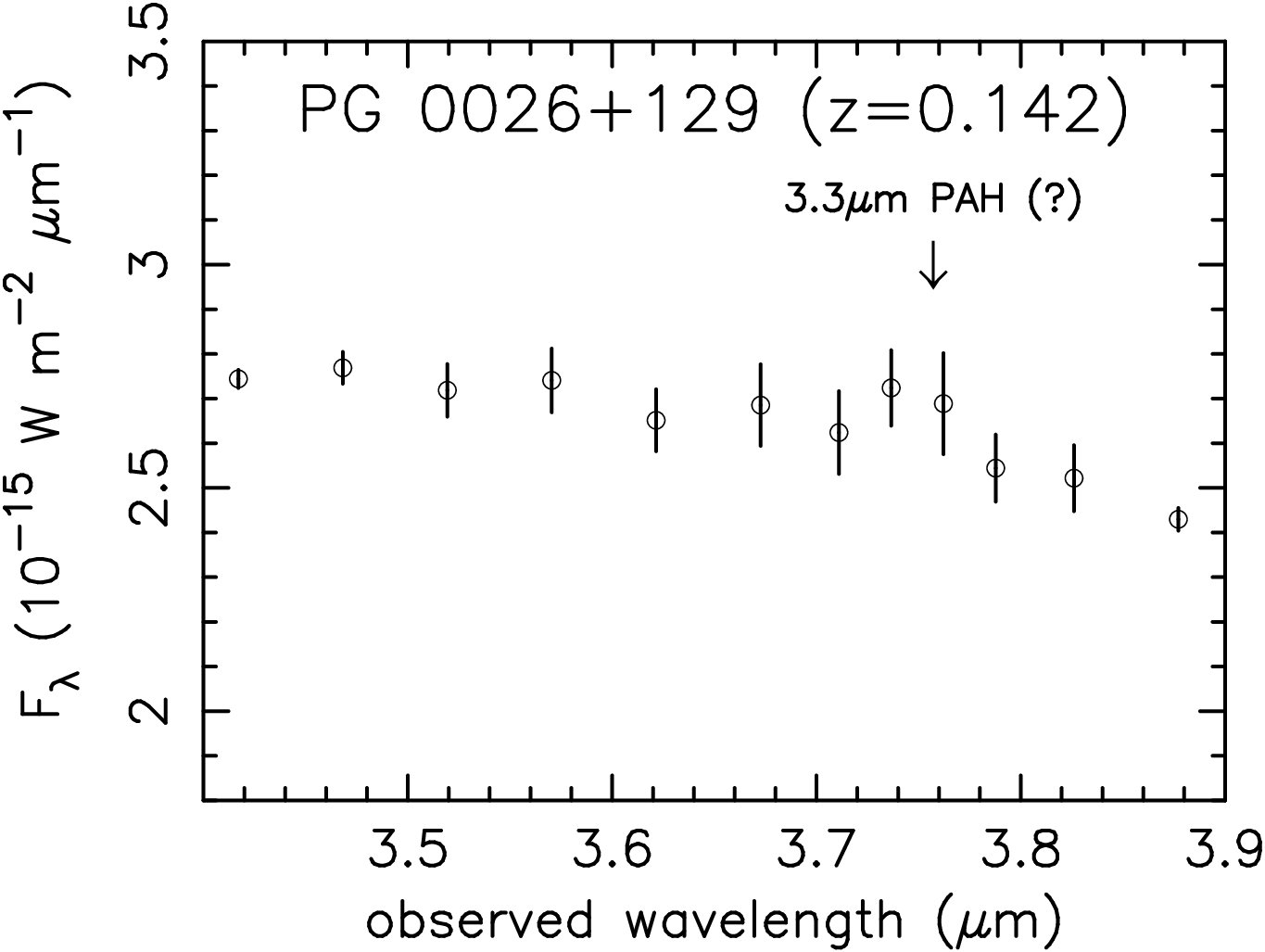} 
\includegraphics[angle=-90,scale=.185]{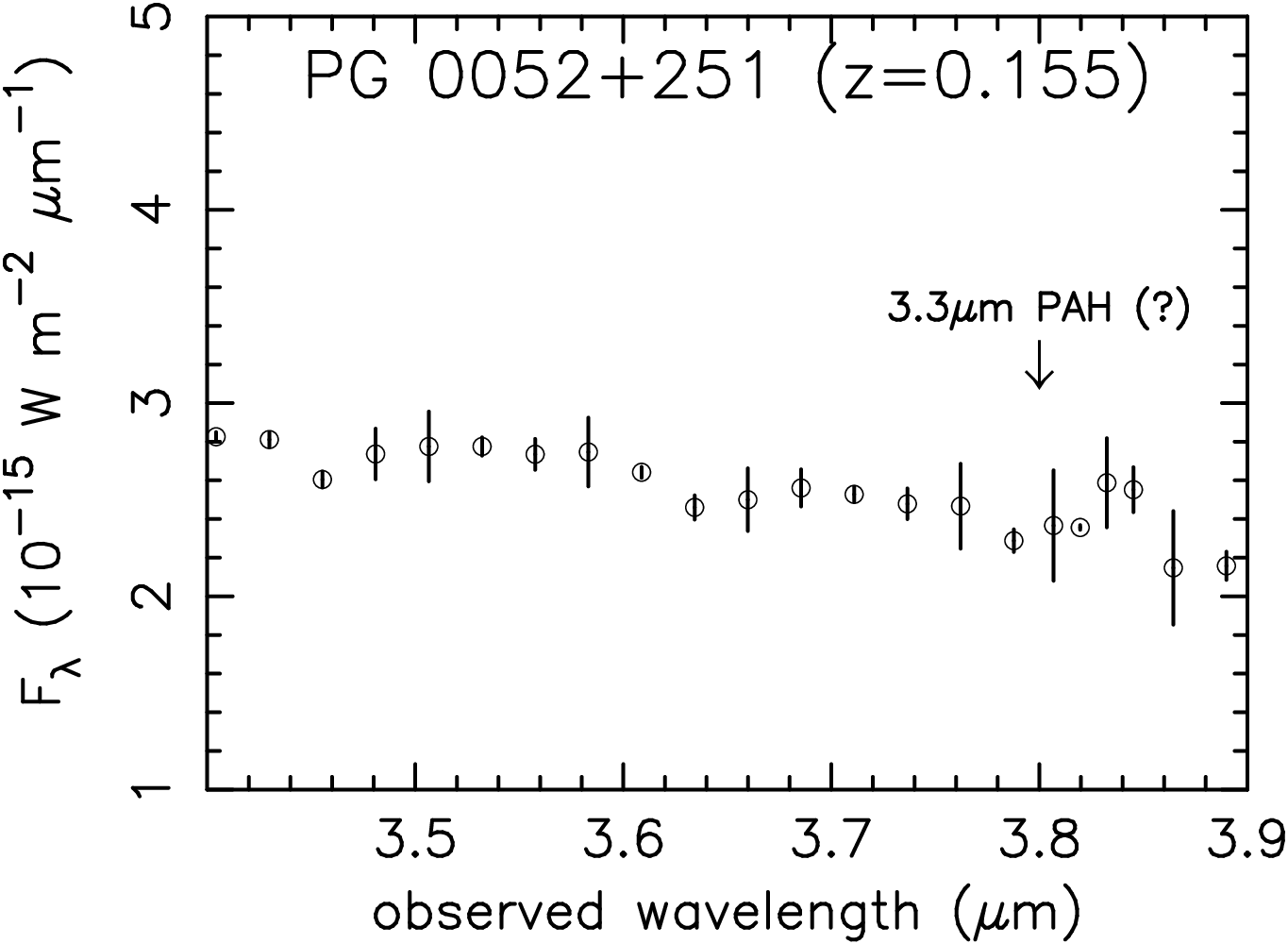} \\ 
\includegraphics[angle=-90,scale=.185]{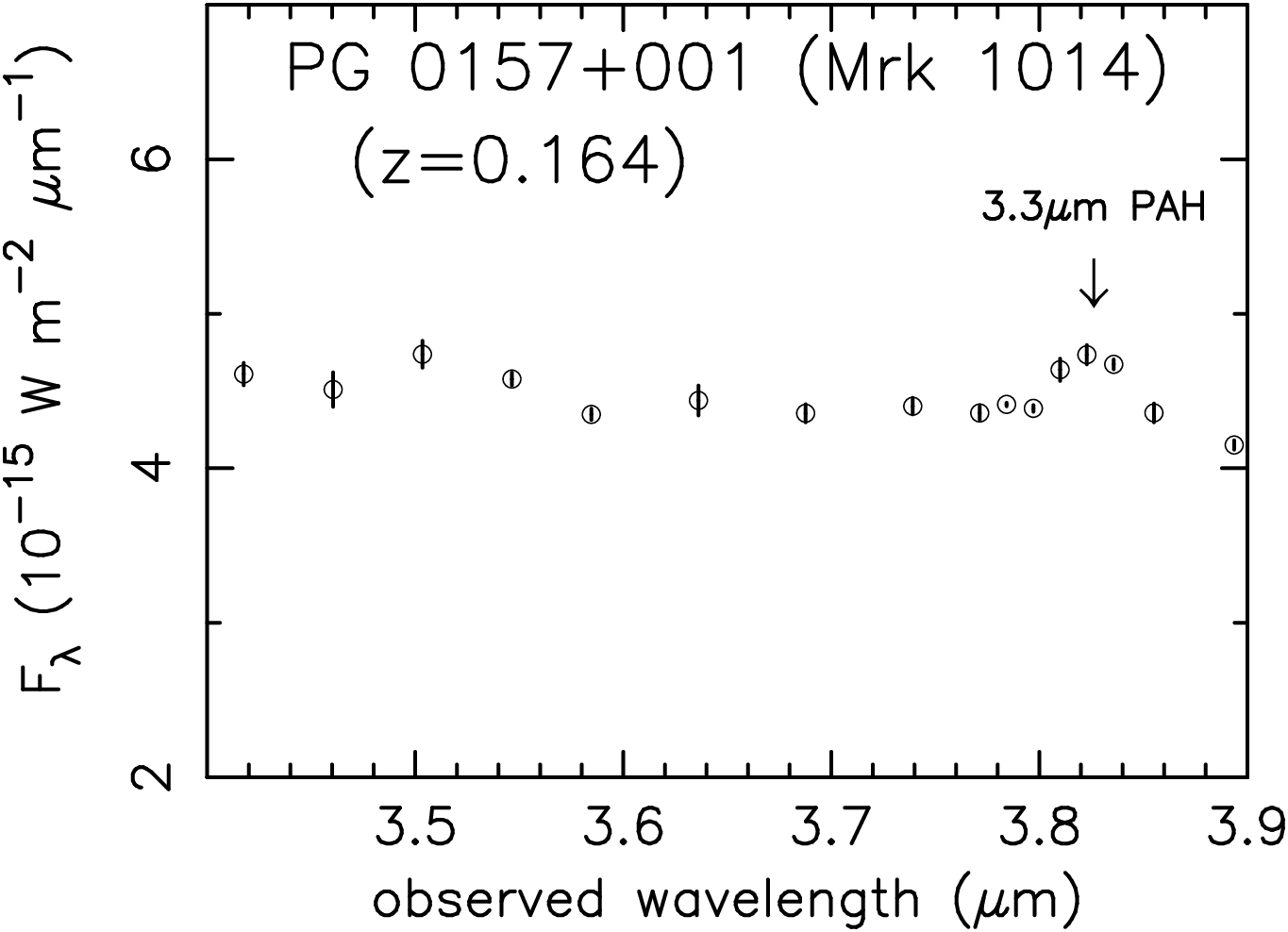} 
\includegraphics[angle=-90,scale=.185]{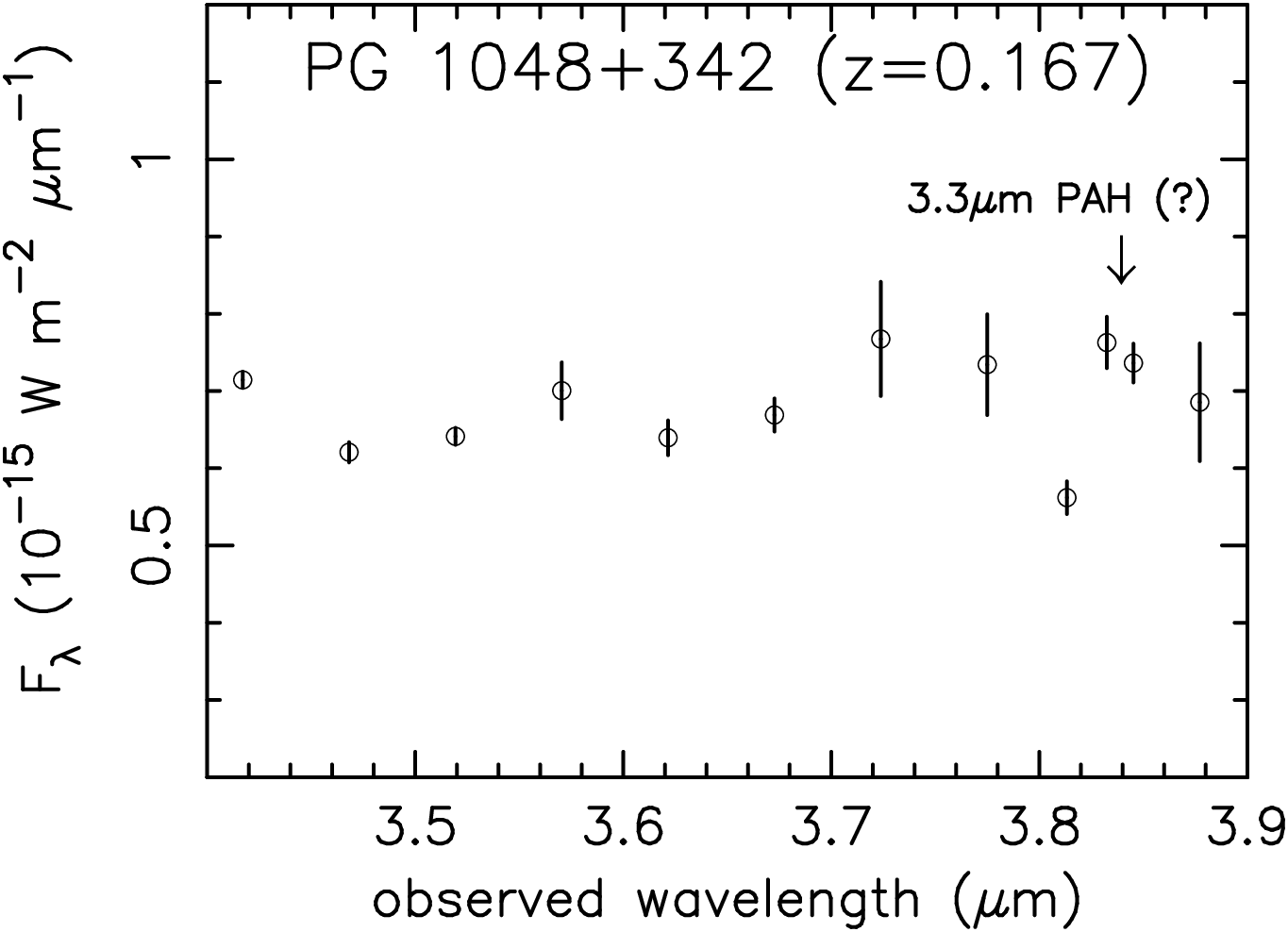} \\
\includegraphics[angle=-90,scale=.185]{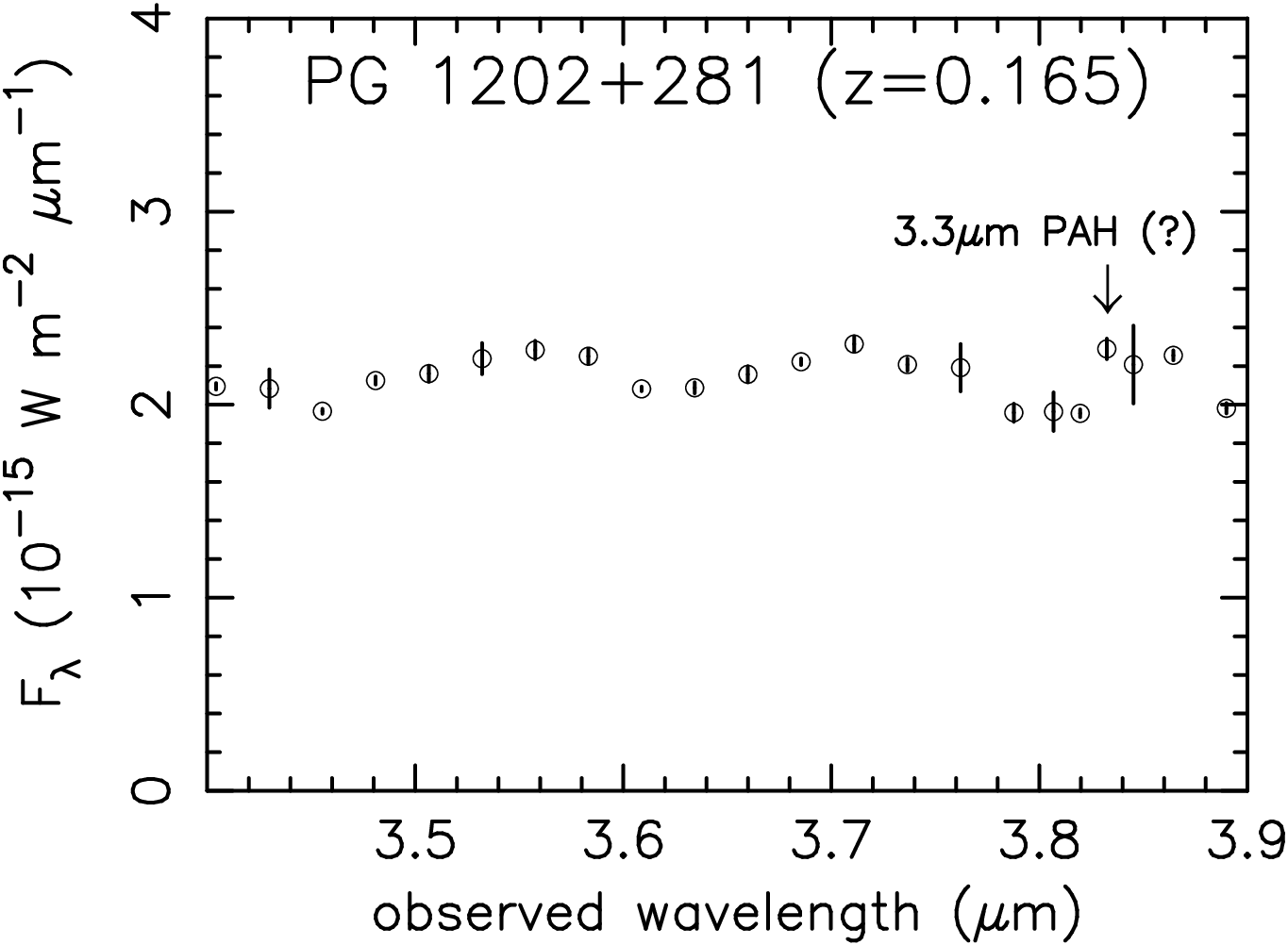} 
\includegraphics[angle=-90,scale=.185]{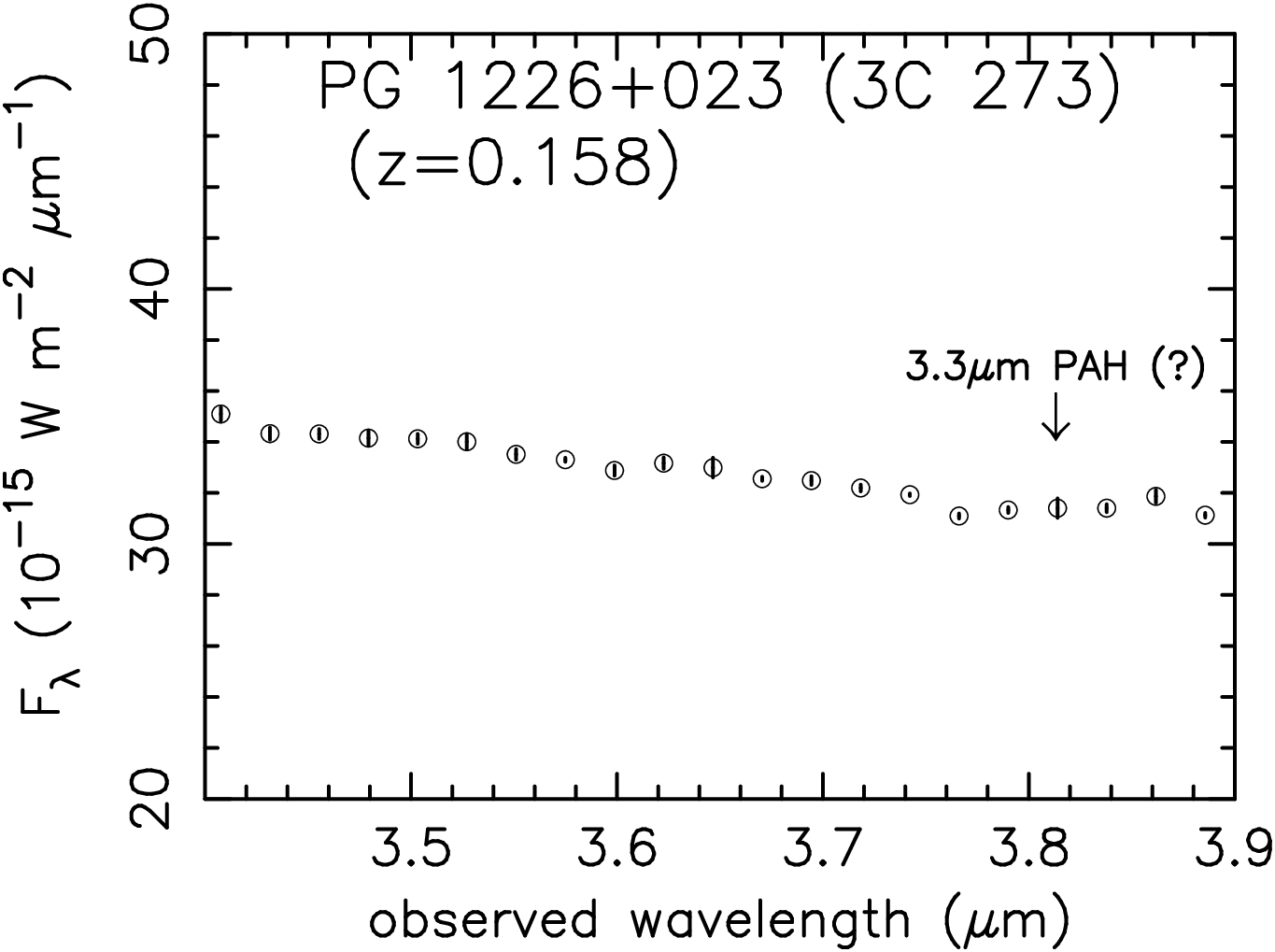} \\
\includegraphics[angle=-90,scale=.185]{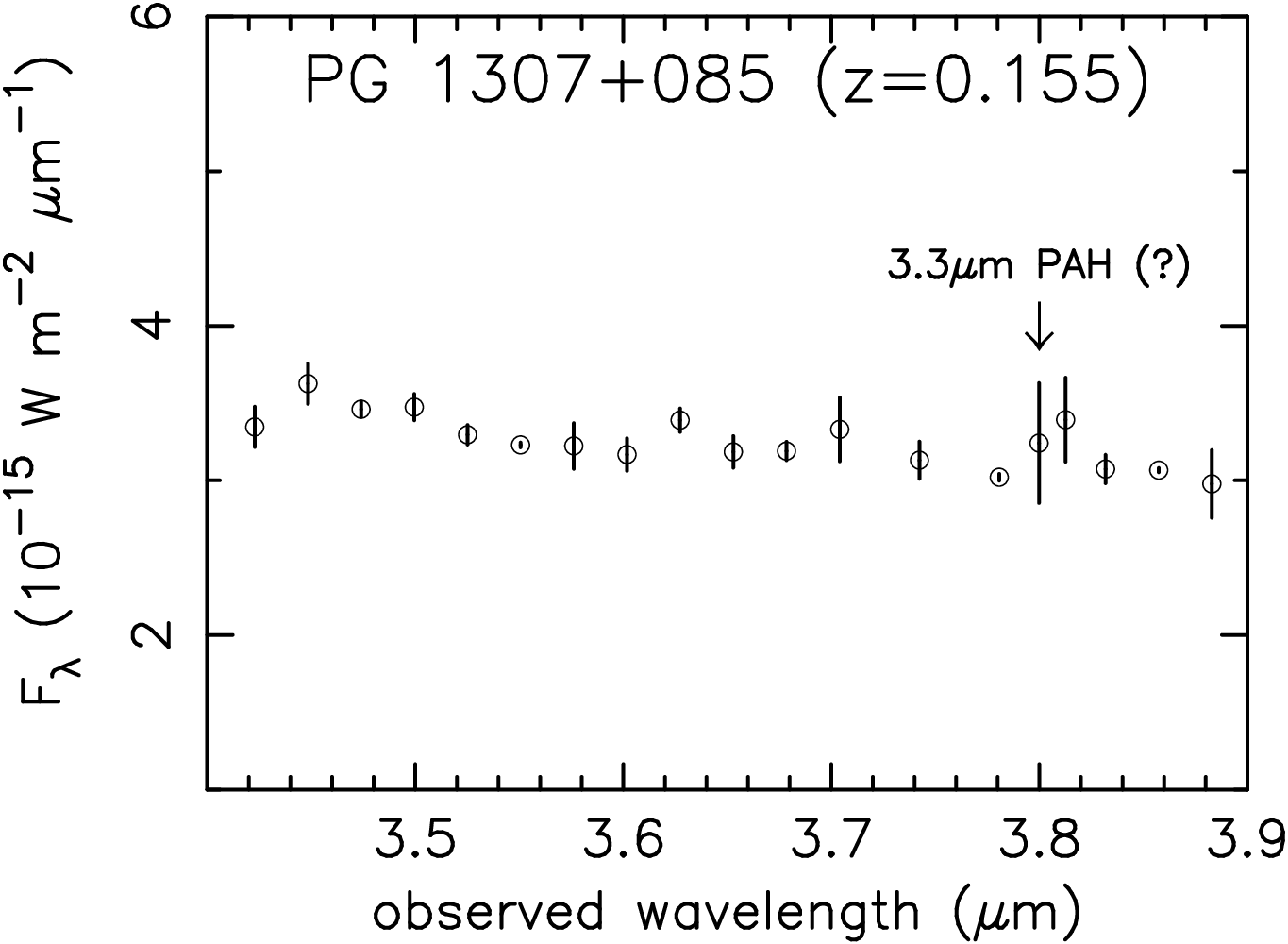} 
\includegraphics[angle=-90,scale=.185]{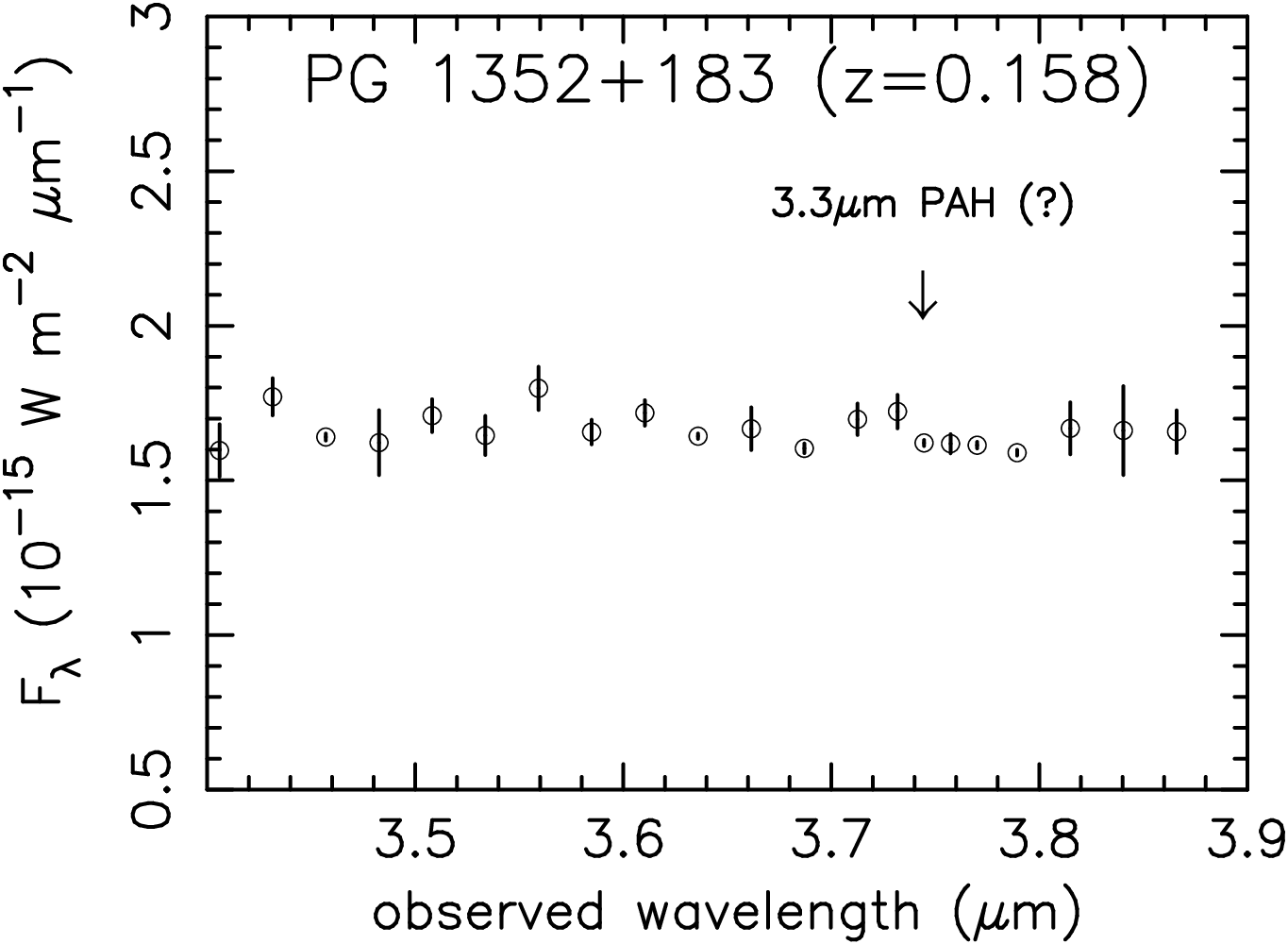} \\
\includegraphics[angle=-90,scale=.185]{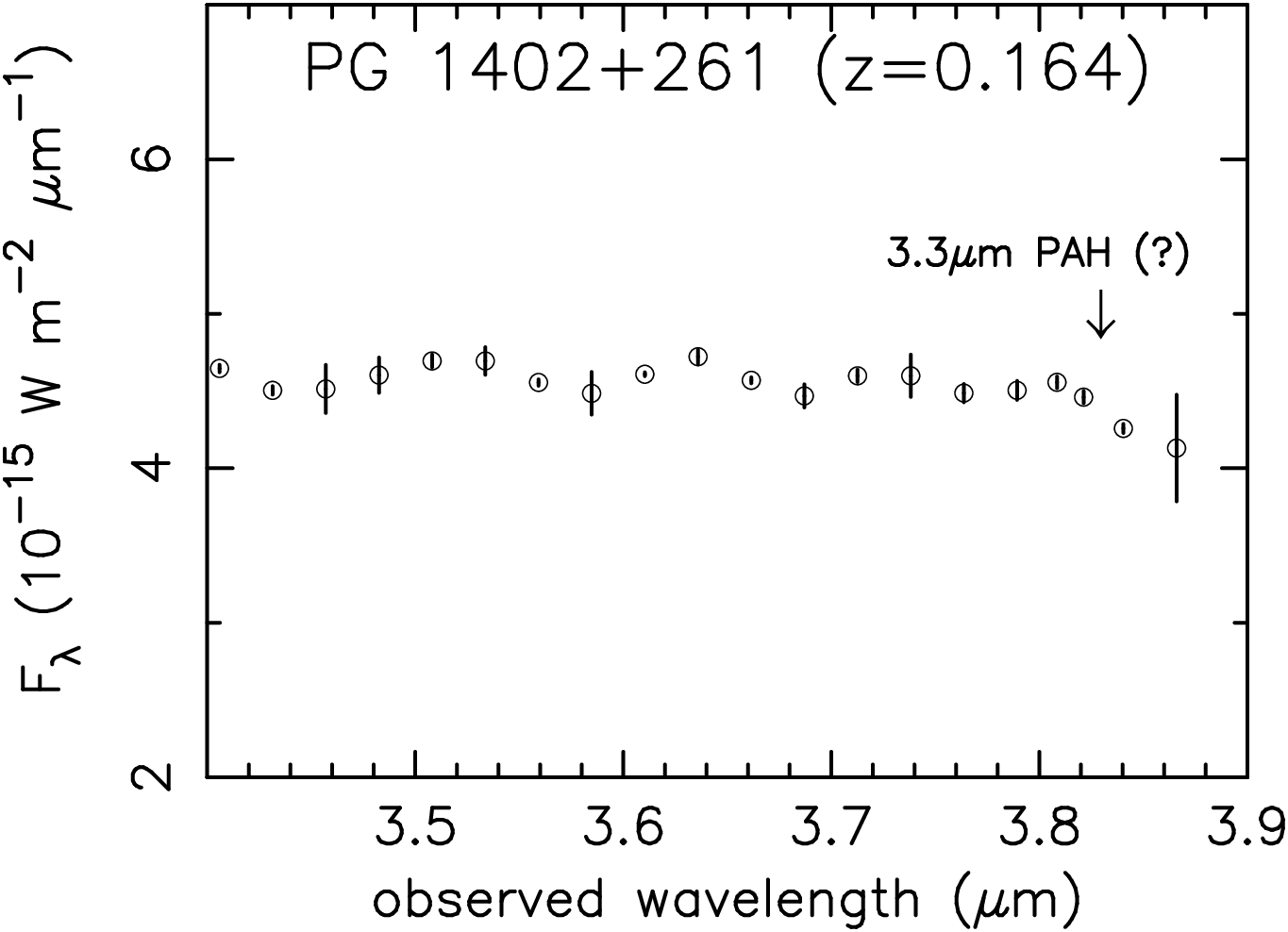} 
\includegraphics[angle=-90,scale=.185]{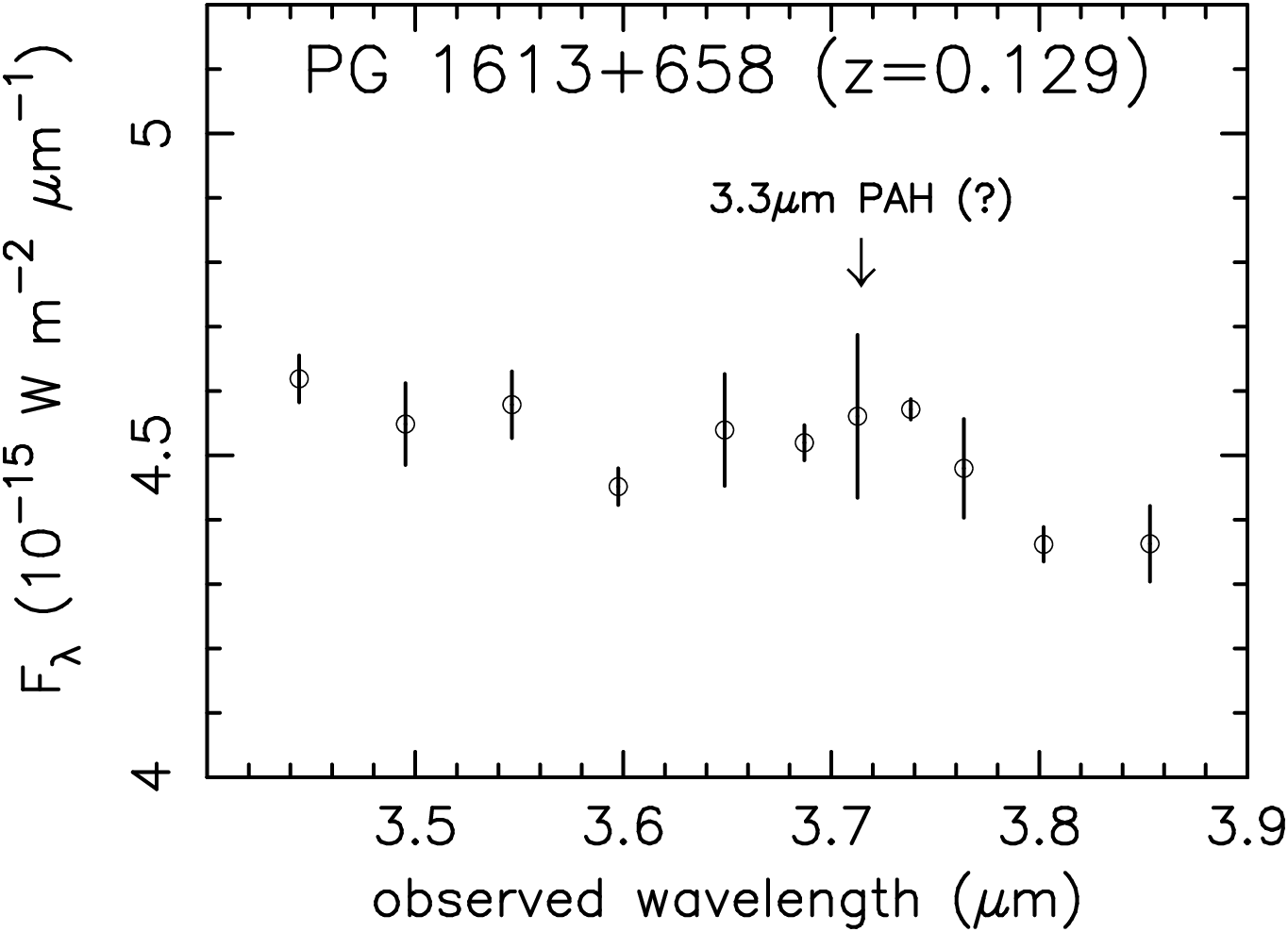} 
\caption{
Infrared 3.4--3.9 $\mu$m spectra of high-luminosity PG QSOs.
The abscissa is the observed wavelength in [$\mu$m], and the
ordinate is the flux F$_{\lambda}$ in 10$^{-15}$ [W m$^{-2}$ $\mu$m$^{-1}$].
The expected wavelength of the 3.3 $\mu$m PAH emission feature is shown.
The "?" indicates that detection is unclear.
}
\end{figure}

%---  Figure 2 ---%
\begin{figure}
\includegraphics[angle=-90,scale=.185]{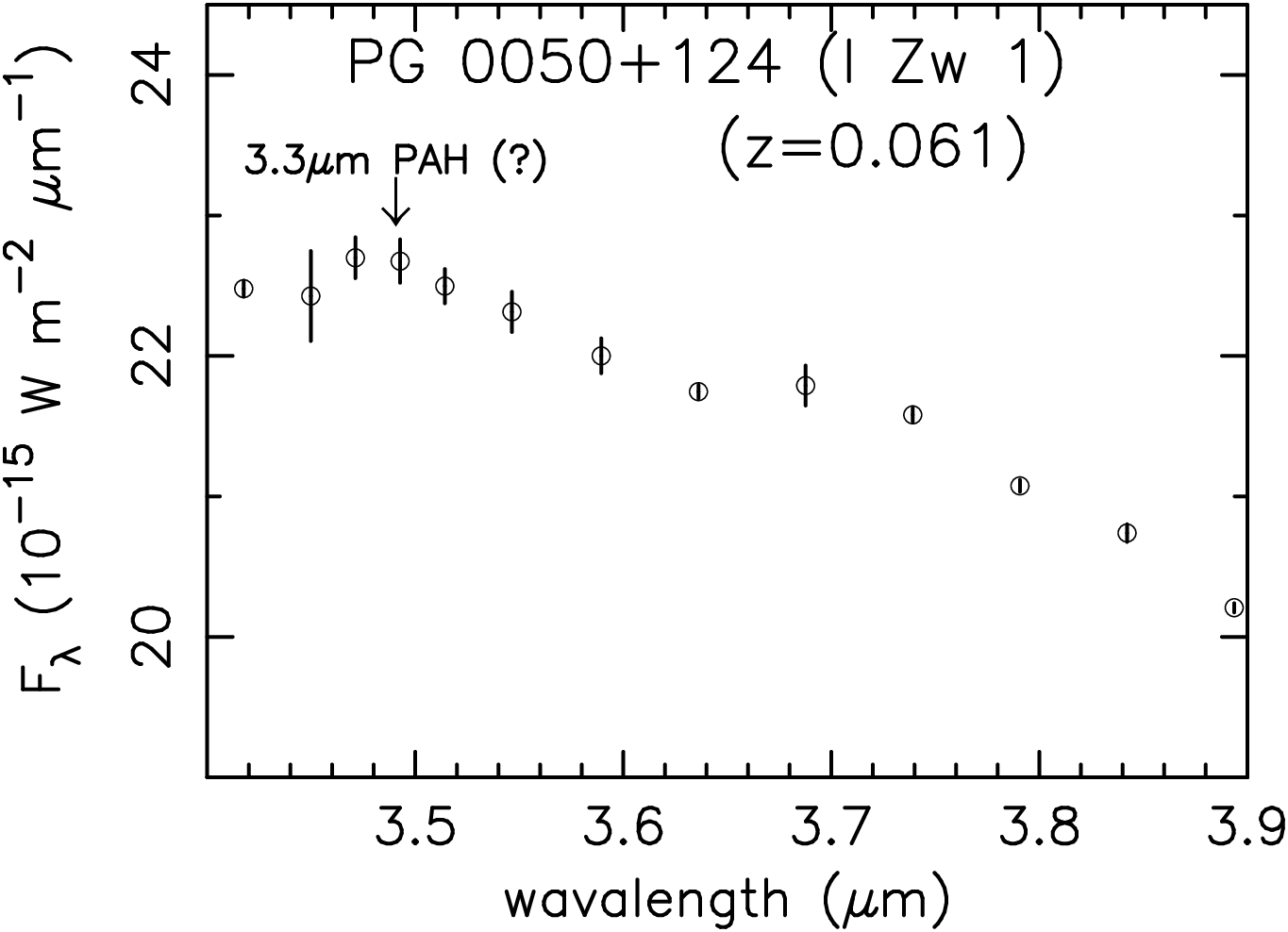} 
\includegraphics[angle=-90,scale=.185]{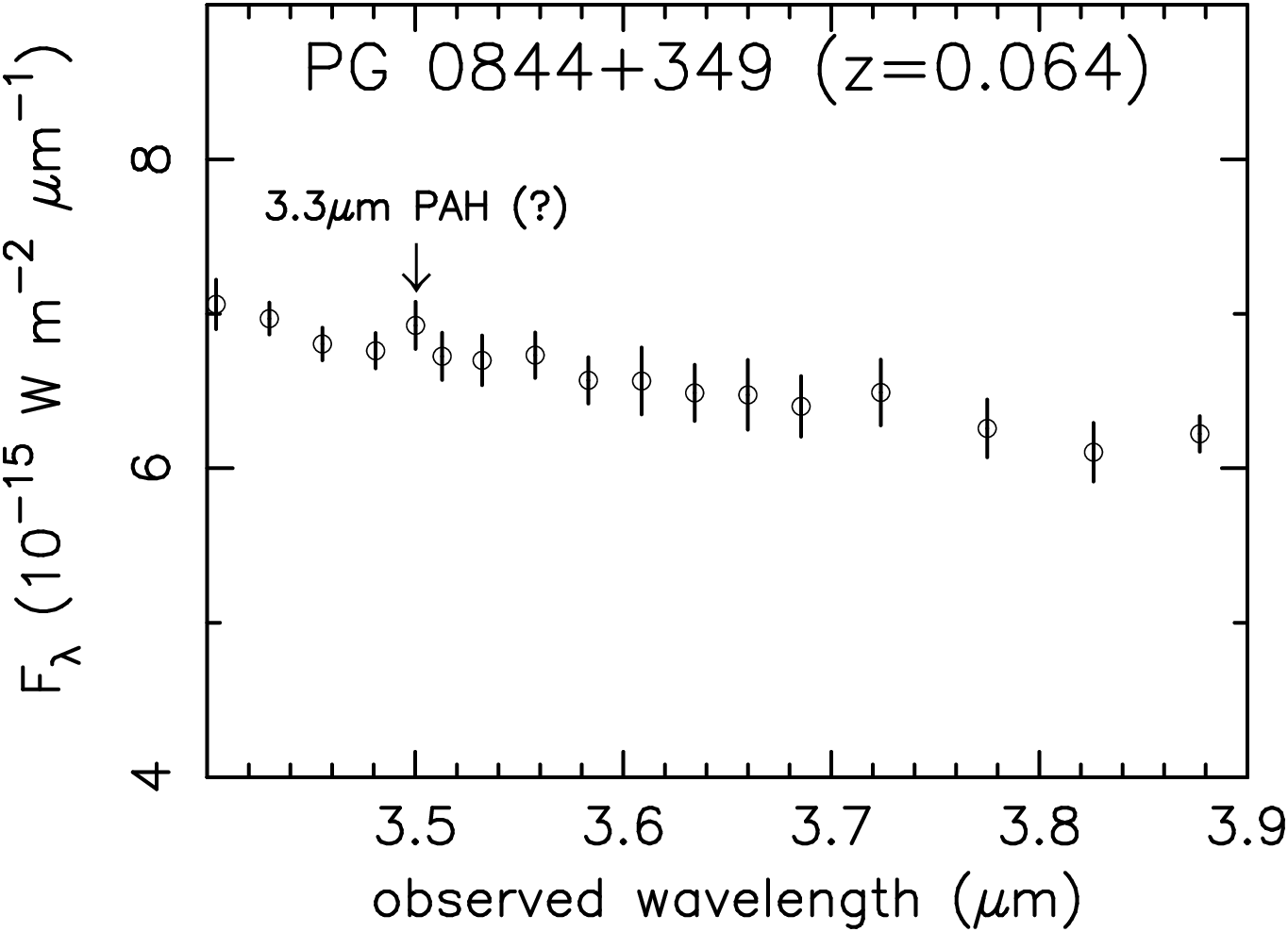} \\ 
\includegraphics[angle=-90,scale=.185]{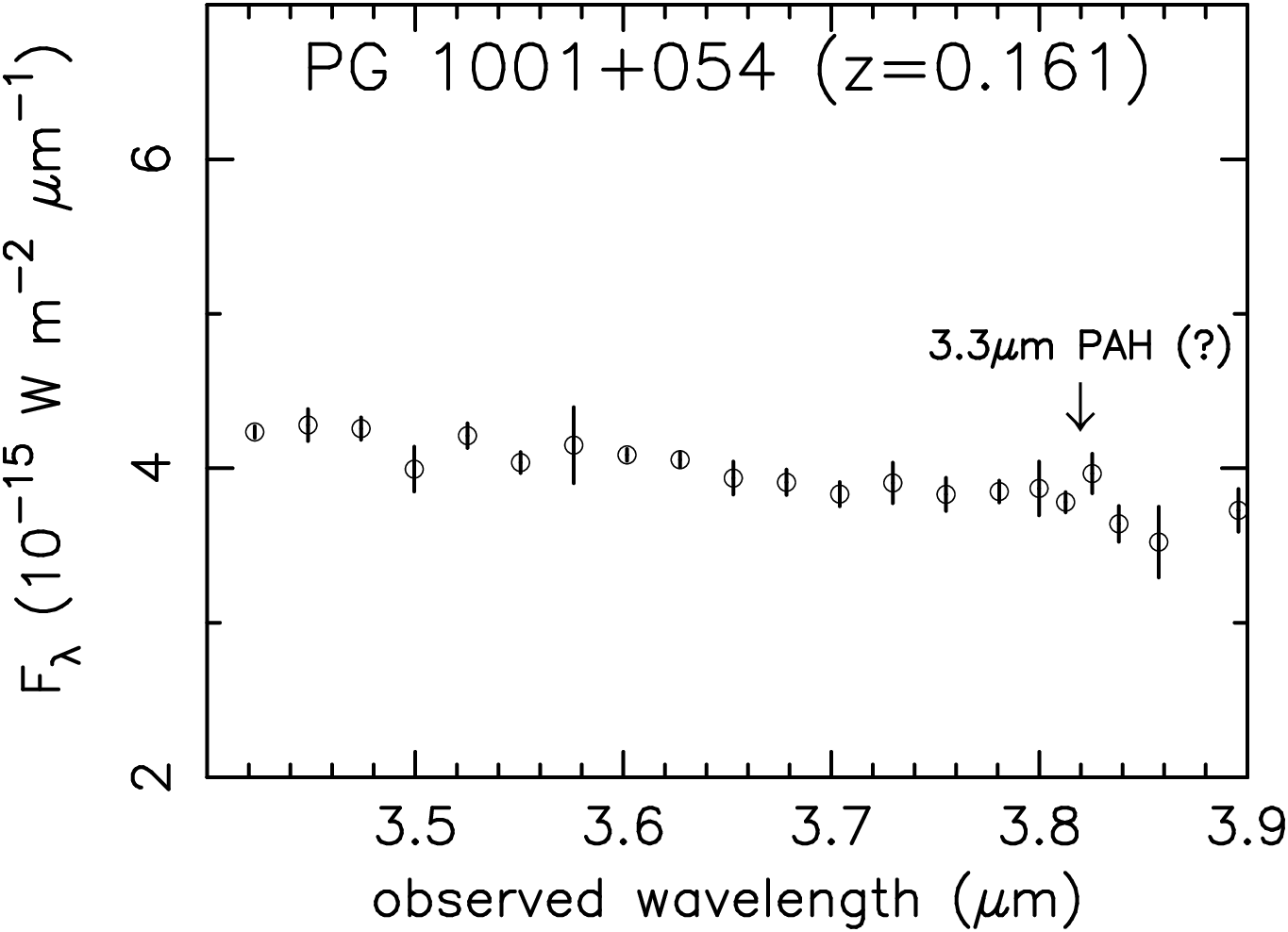} 
\includegraphics[angle=-90,scale=.185]{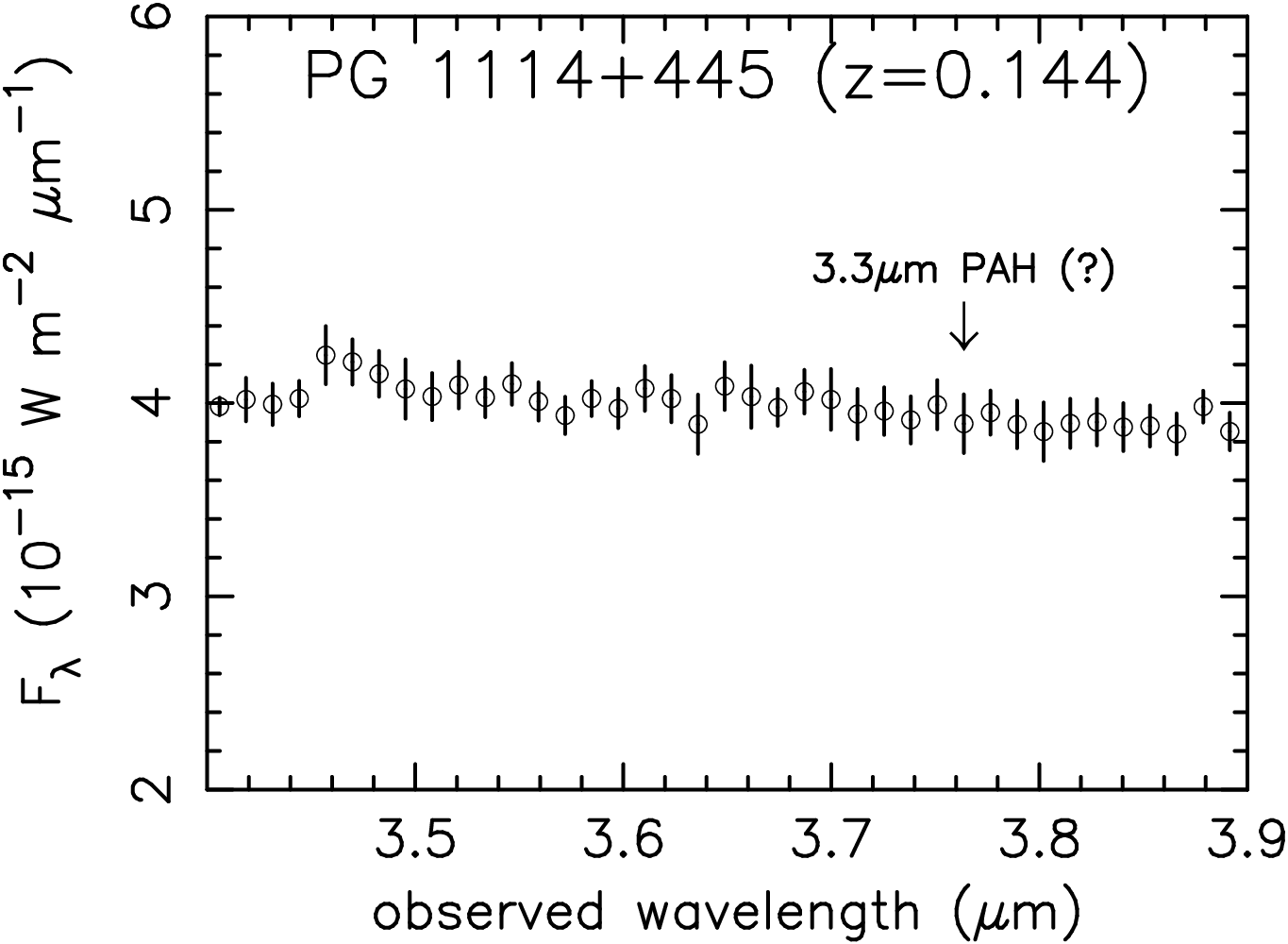} \\
\includegraphics[angle=-90,scale=.185]{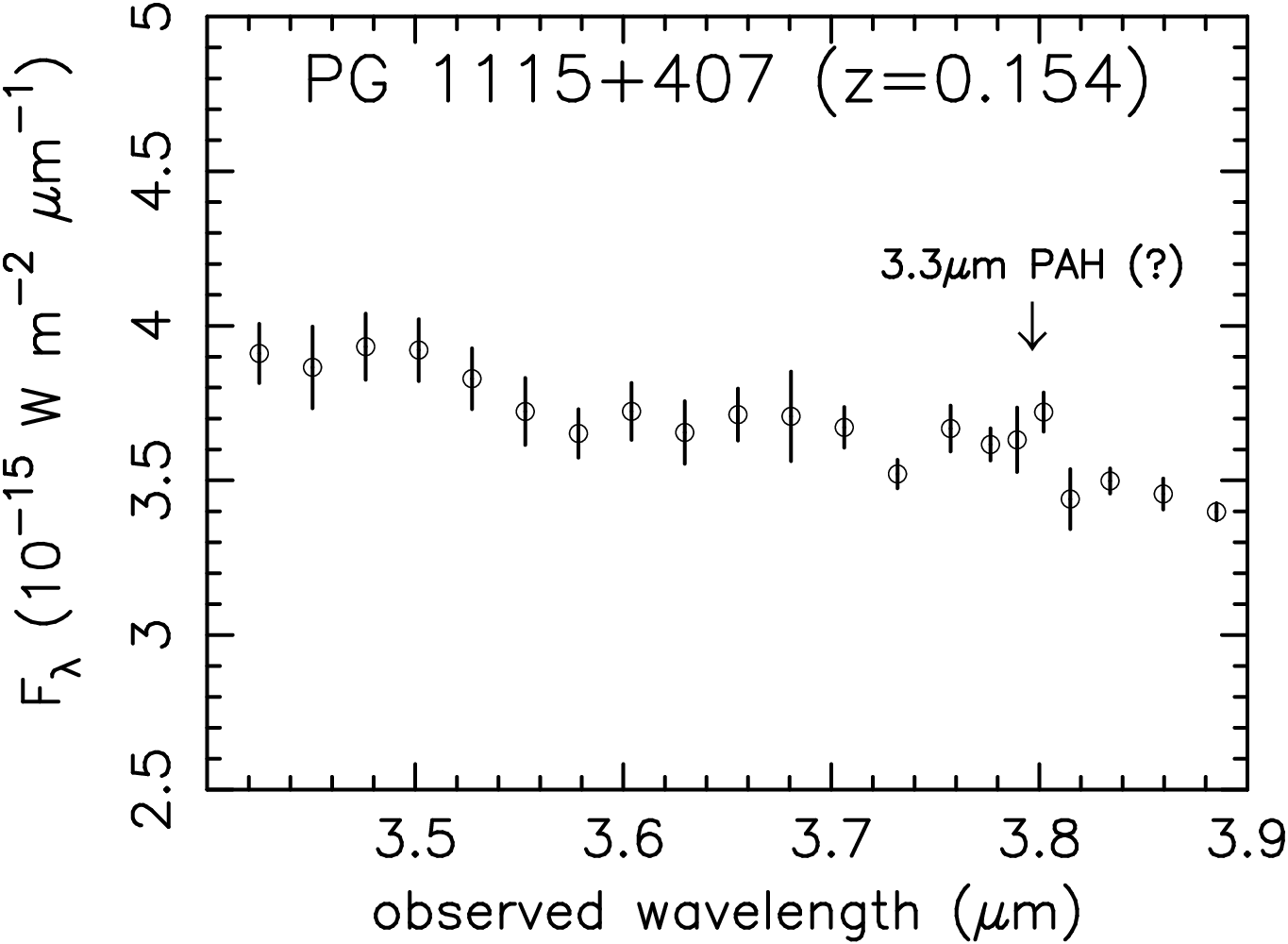} 
\includegraphics[angle=-90,scale=.185]{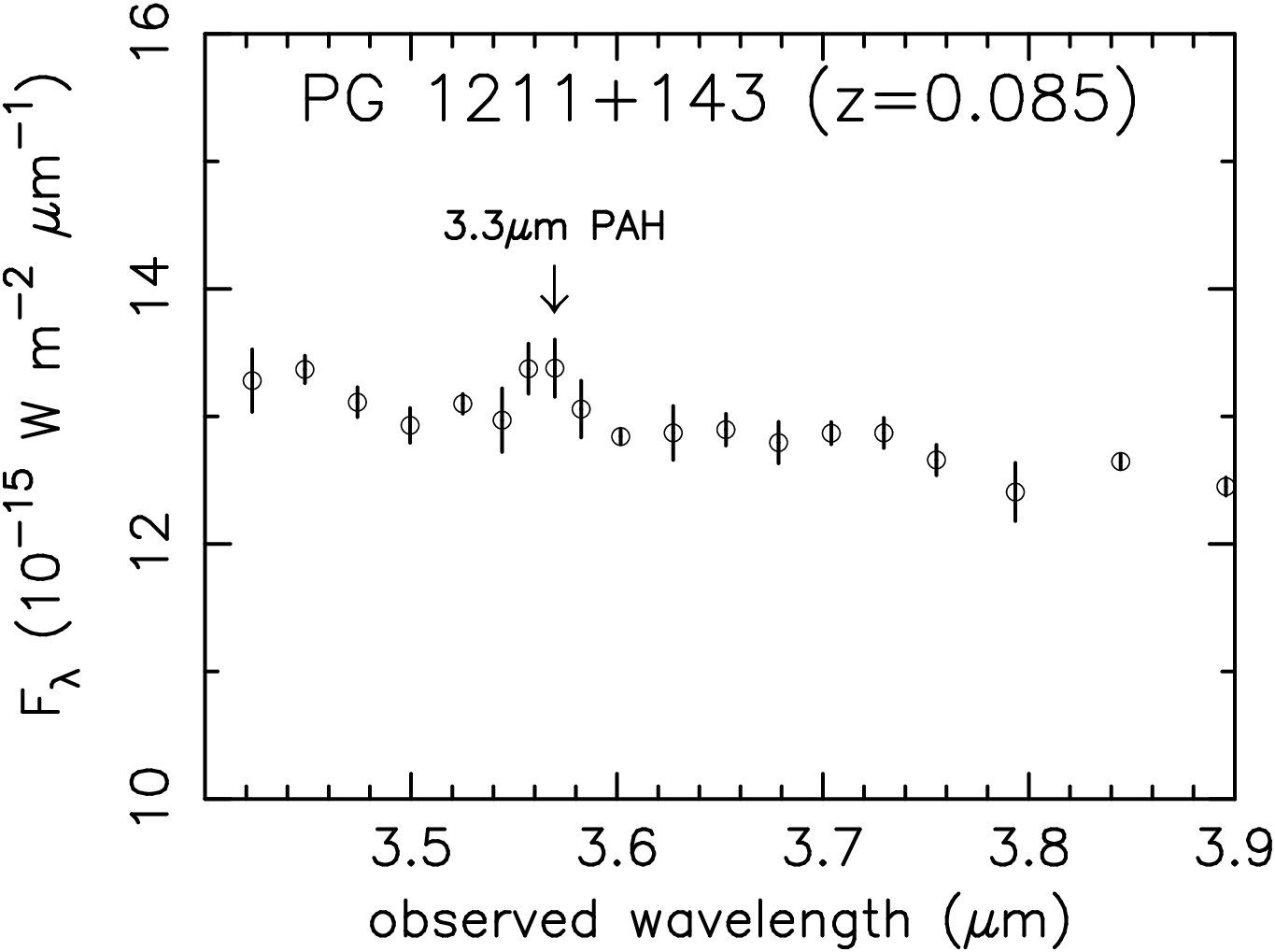} \\
\includegraphics[angle=-90,scale=.185]{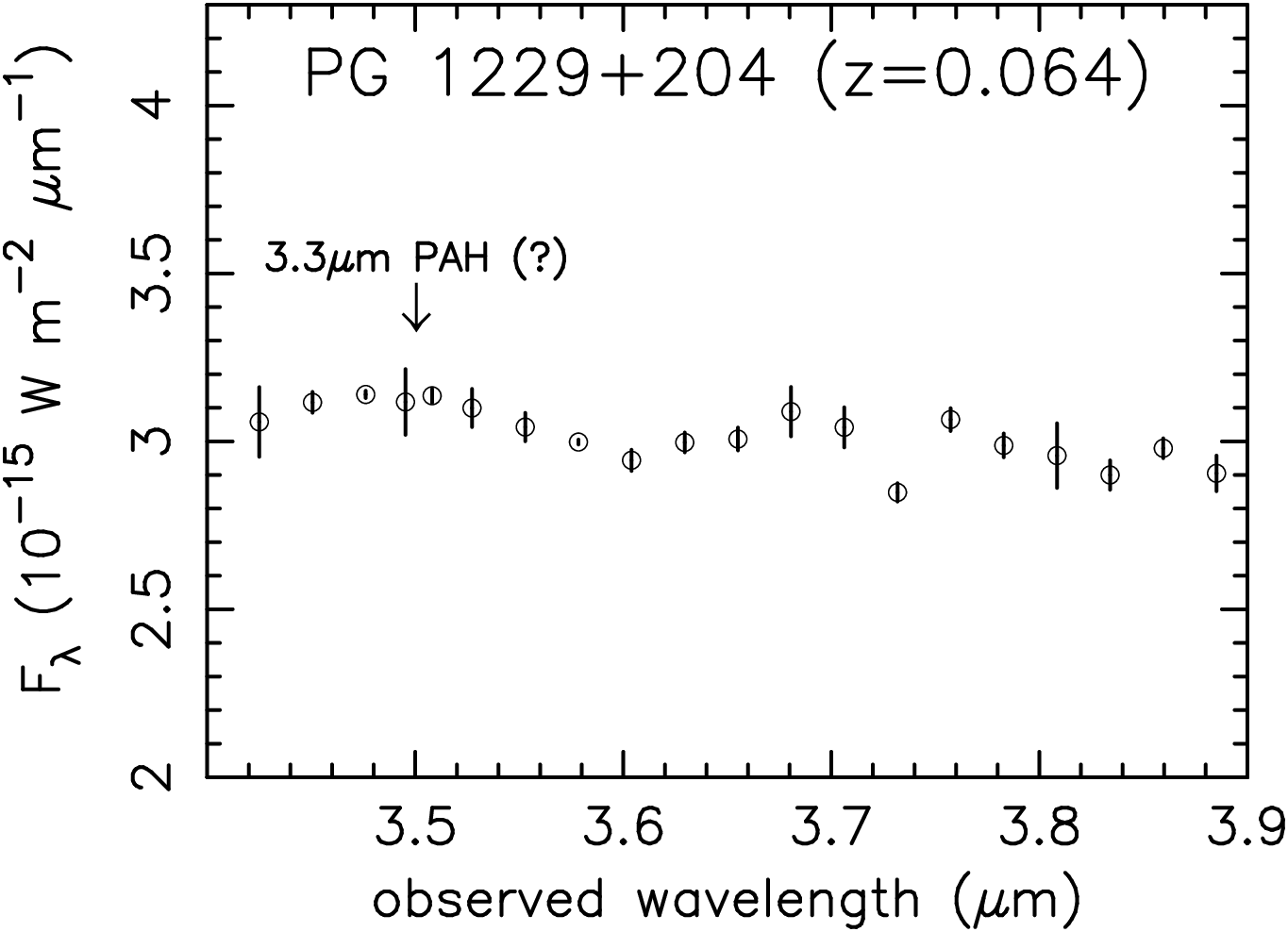} 
\includegraphics[angle=-90,scale=.185]{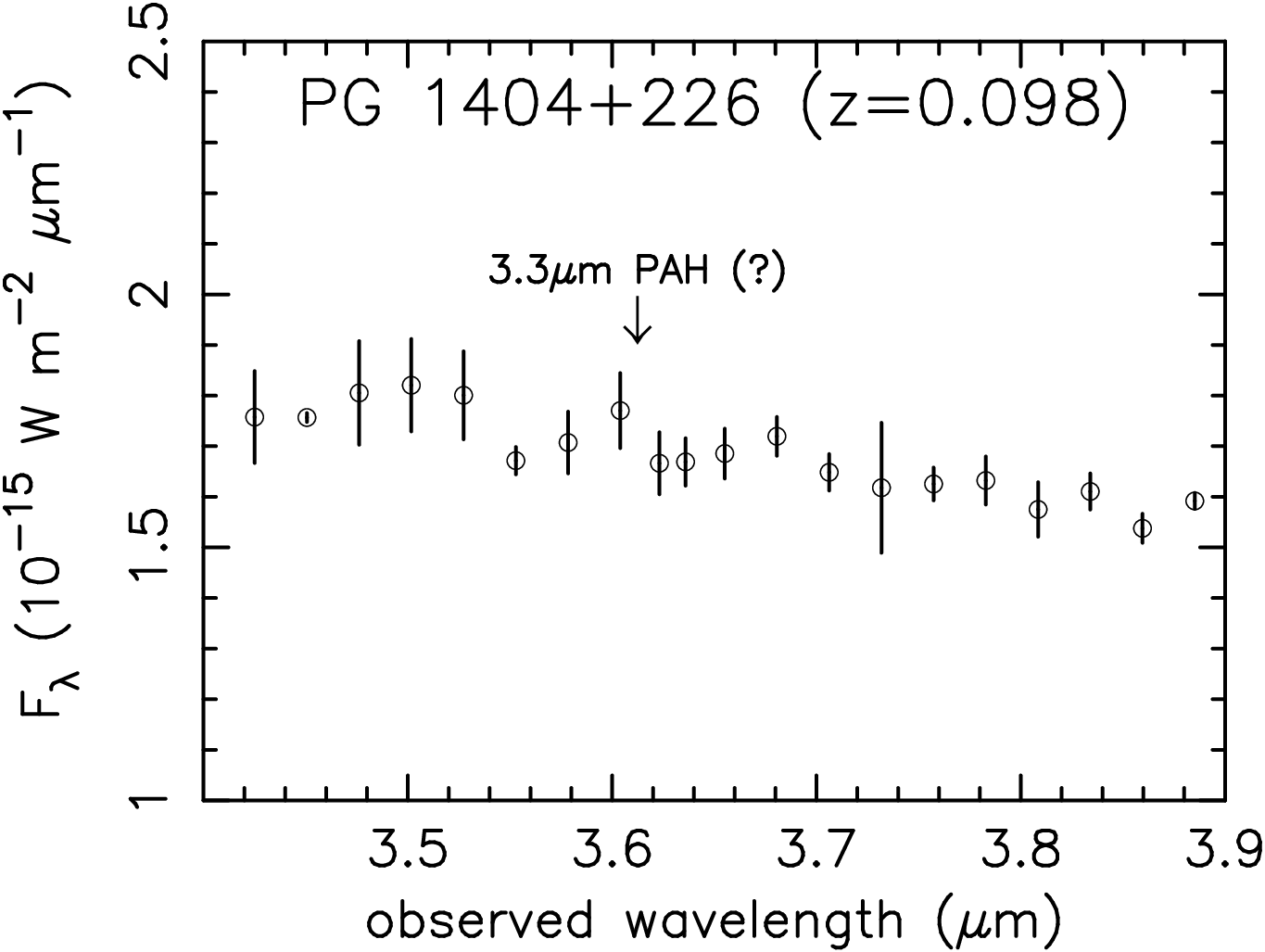} \\
\includegraphics[angle=-90,scale=.185]{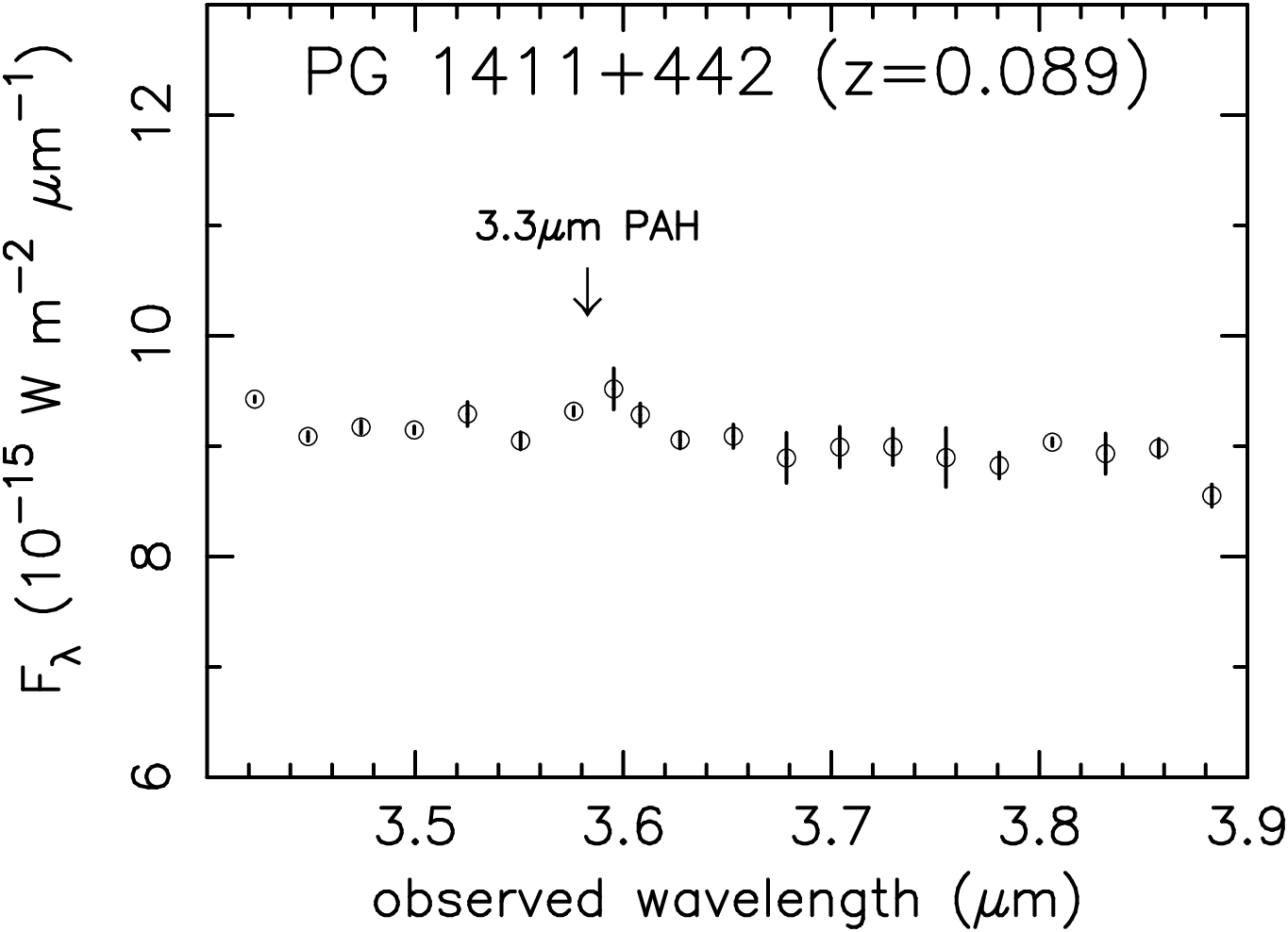} 
\includegraphics[angle=-90,scale=.185]{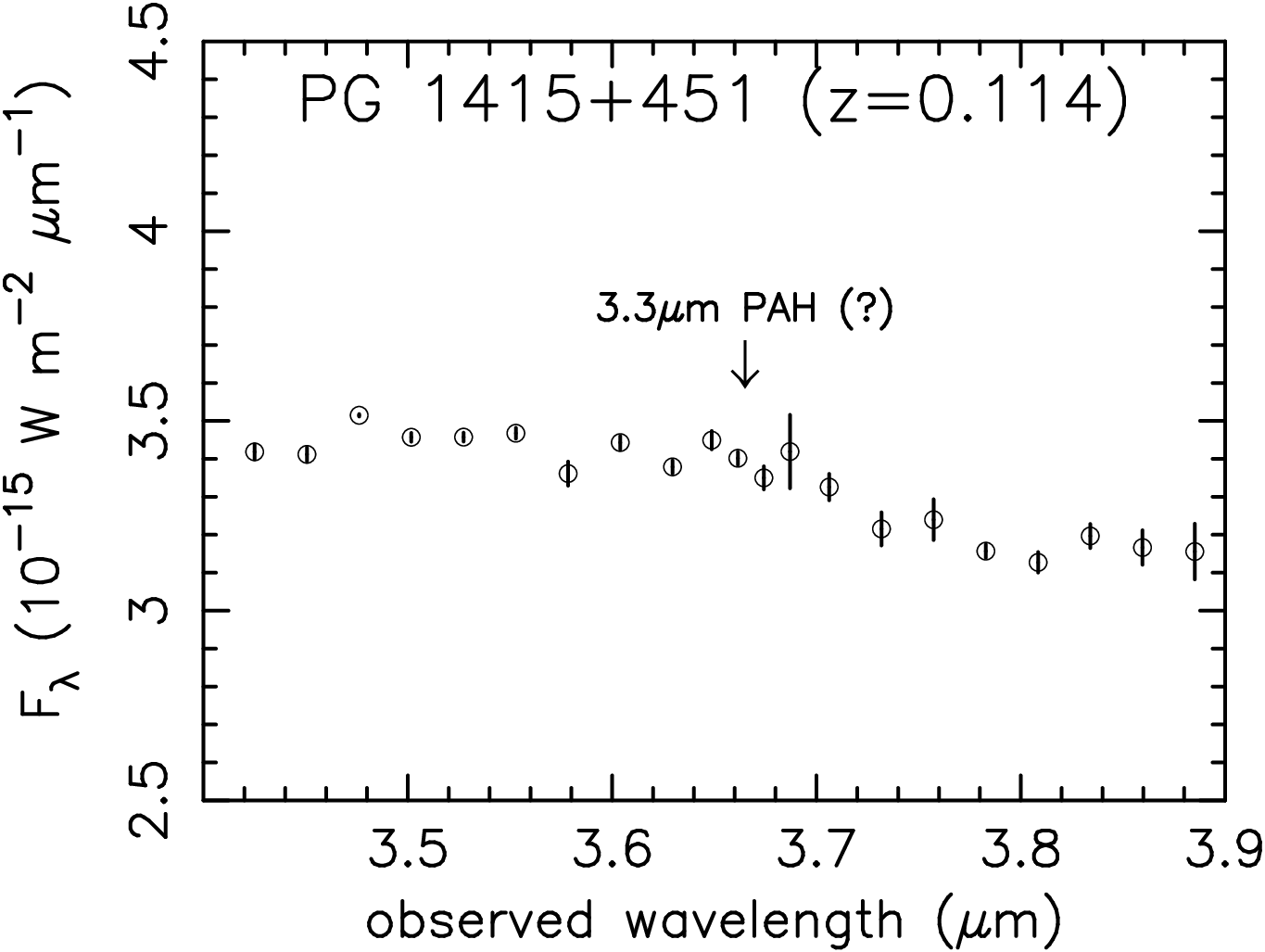} \\
\includegraphics[angle=-90,scale=.185]{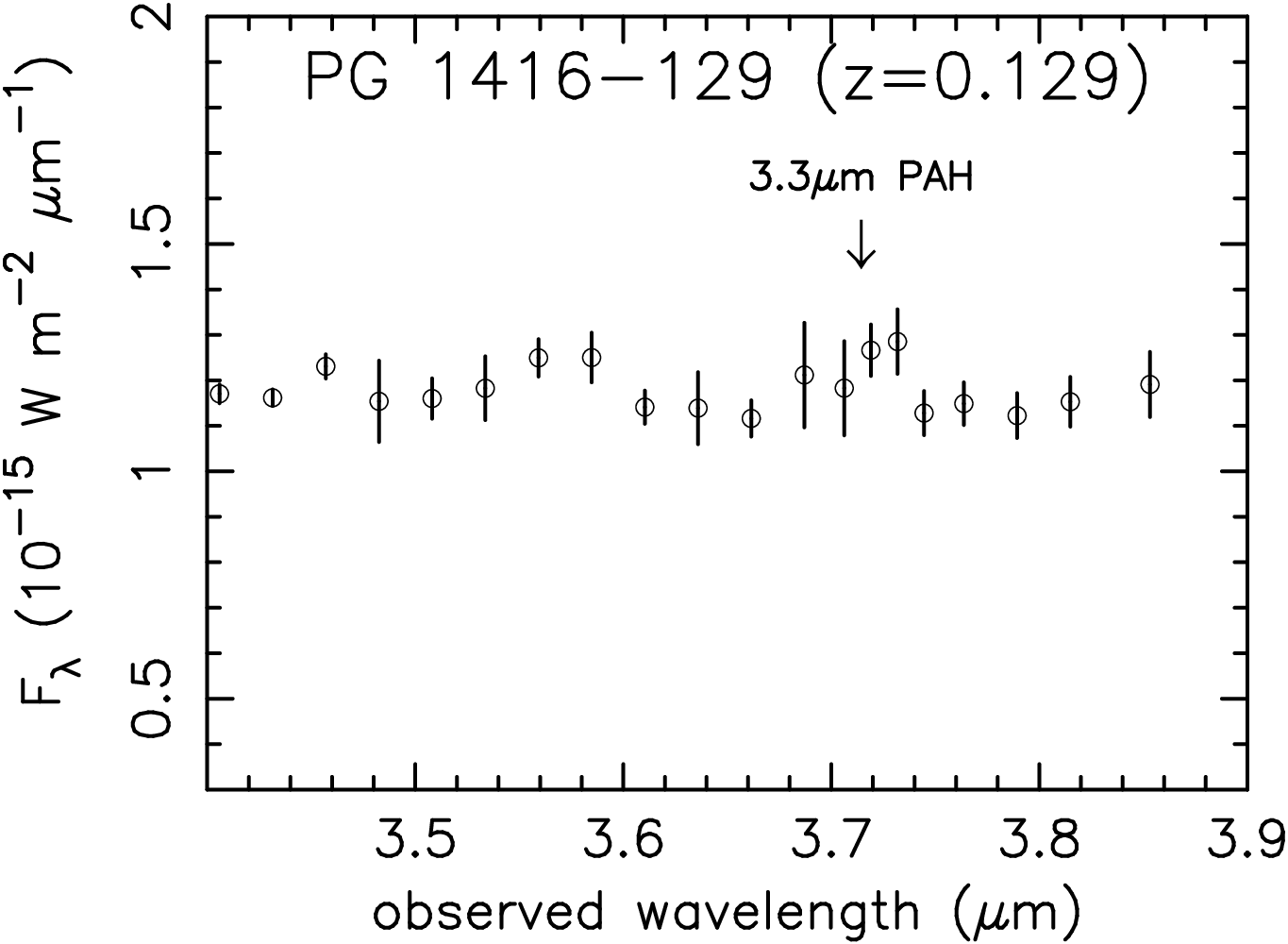} 
\includegraphics[angle=-90,scale=.185]{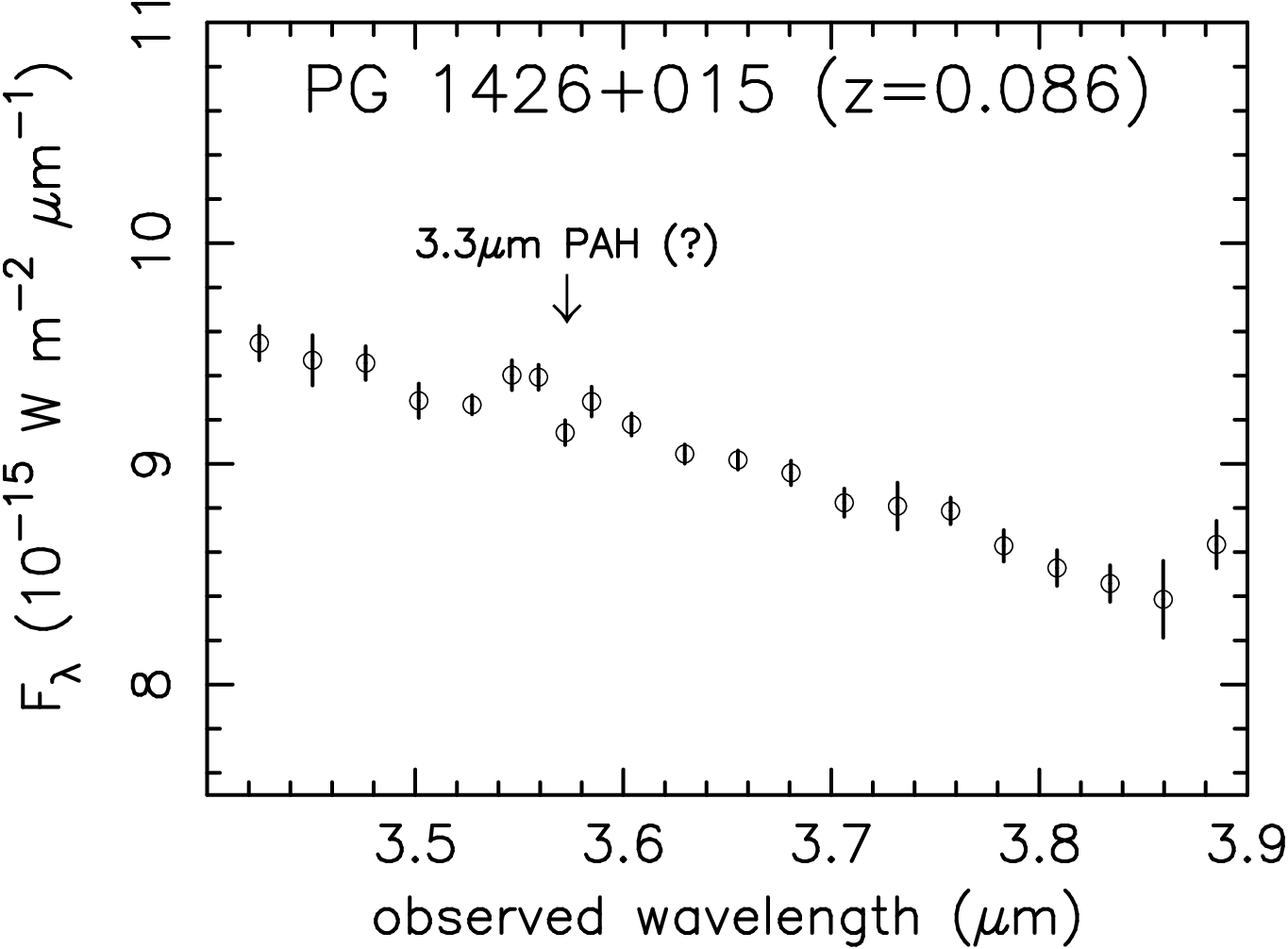} \\
\includegraphics[angle=-90,scale=.185]{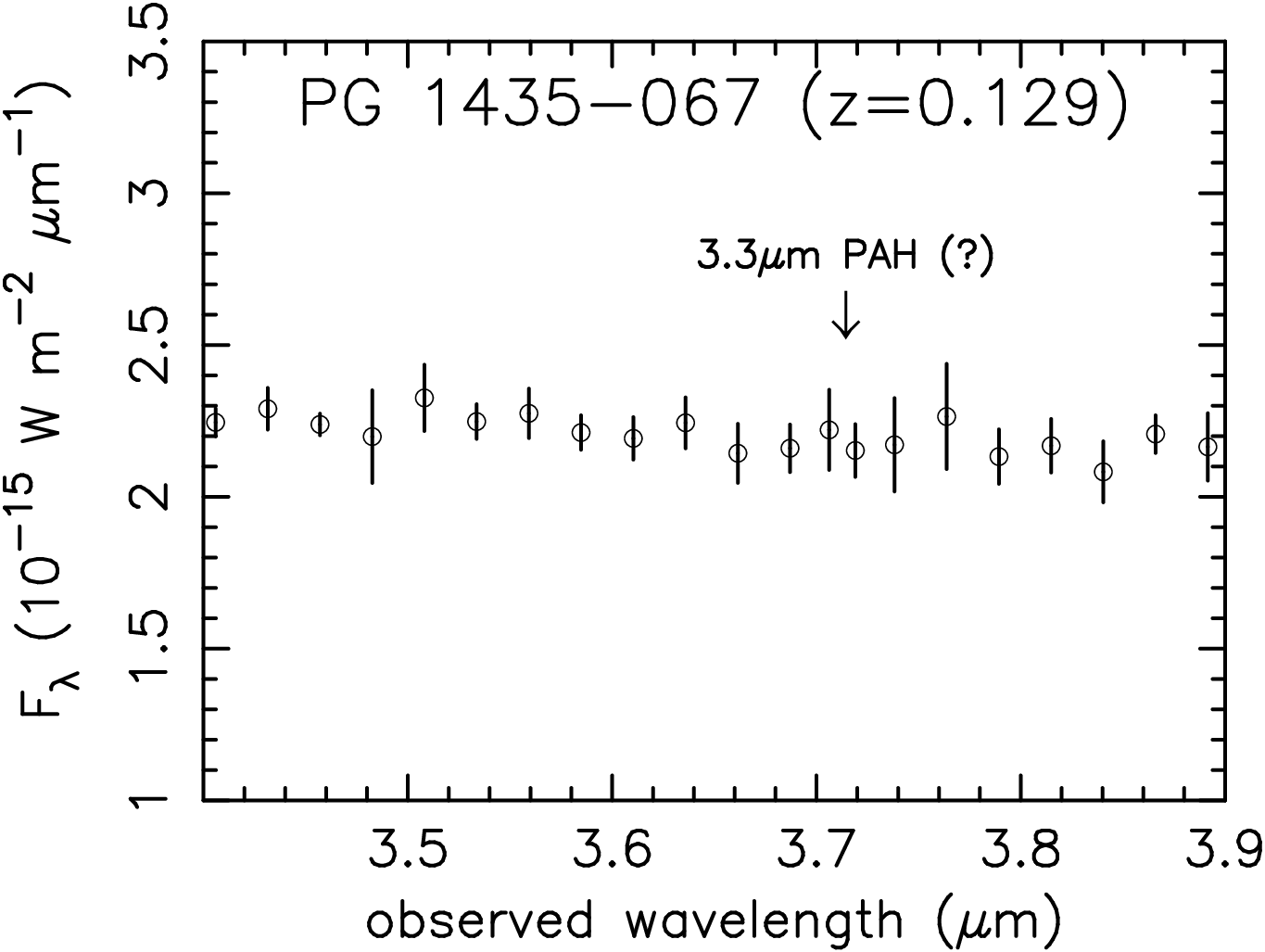} 
\includegraphics[angle=-90,scale=.185]{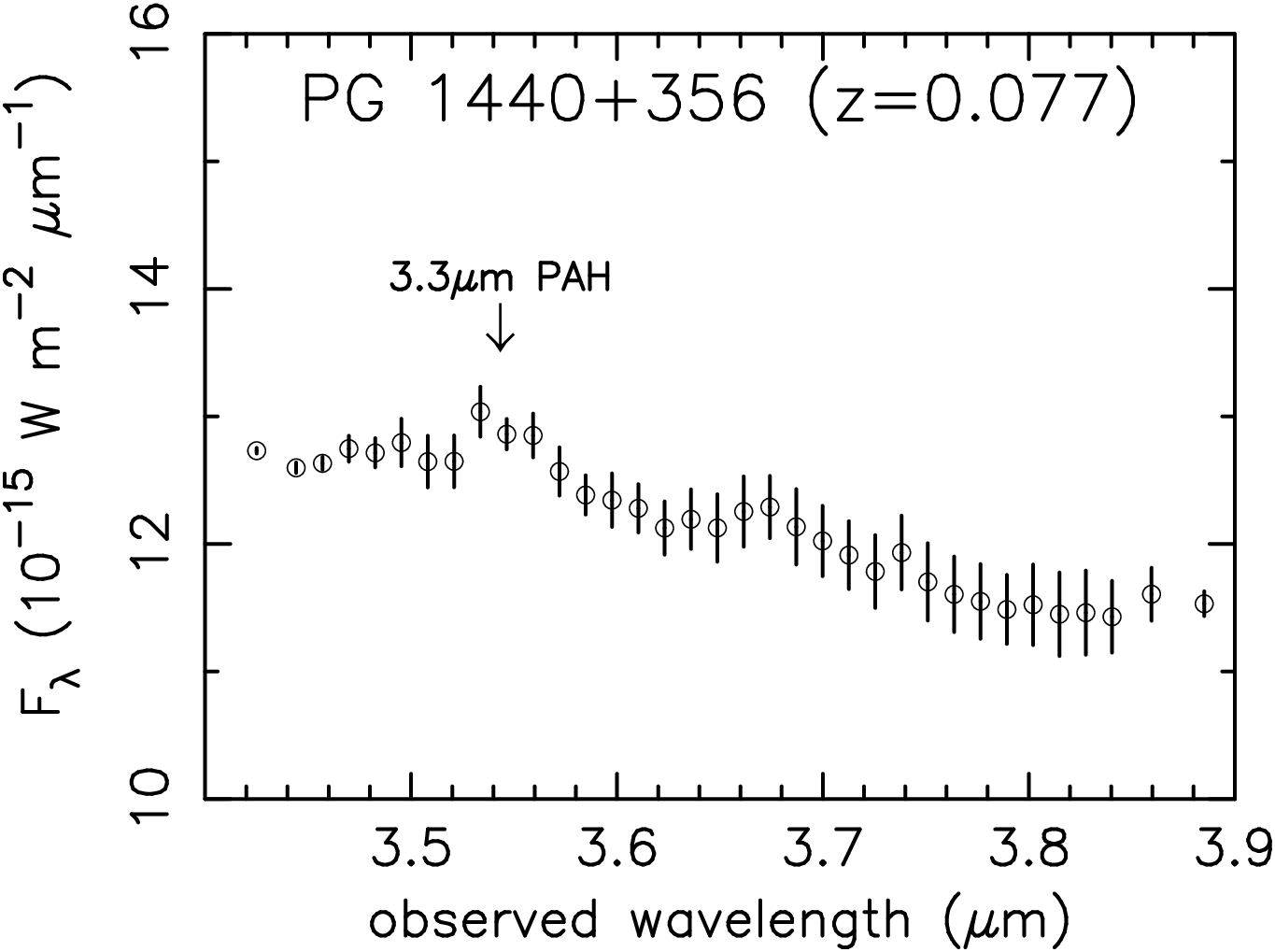} \\
\end{figure}

\clearpage

\begin{figure}
\includegraphics[angle=-90,scale=.185]{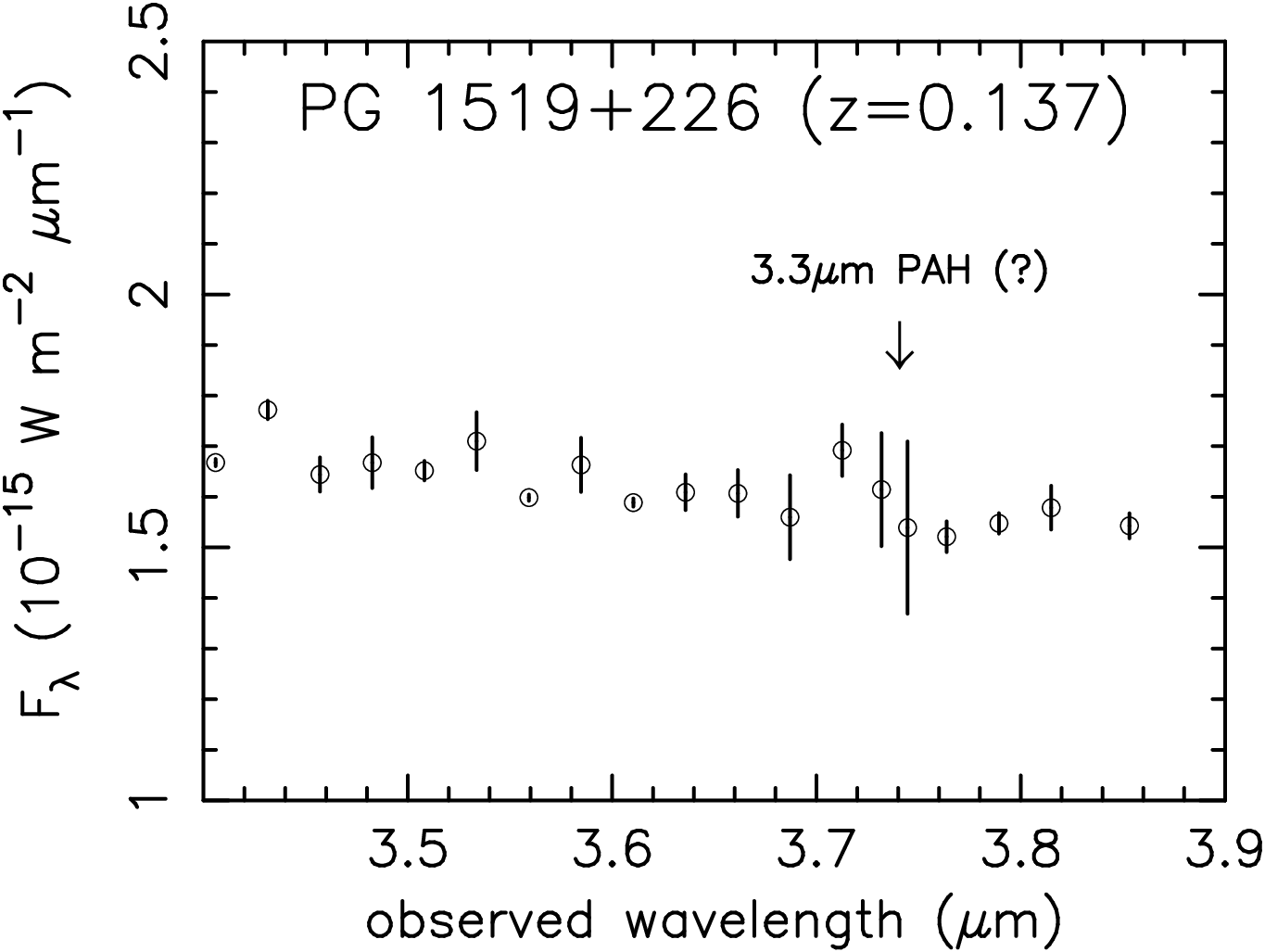} 
\includegraphics[angle=-90,scale=.185]{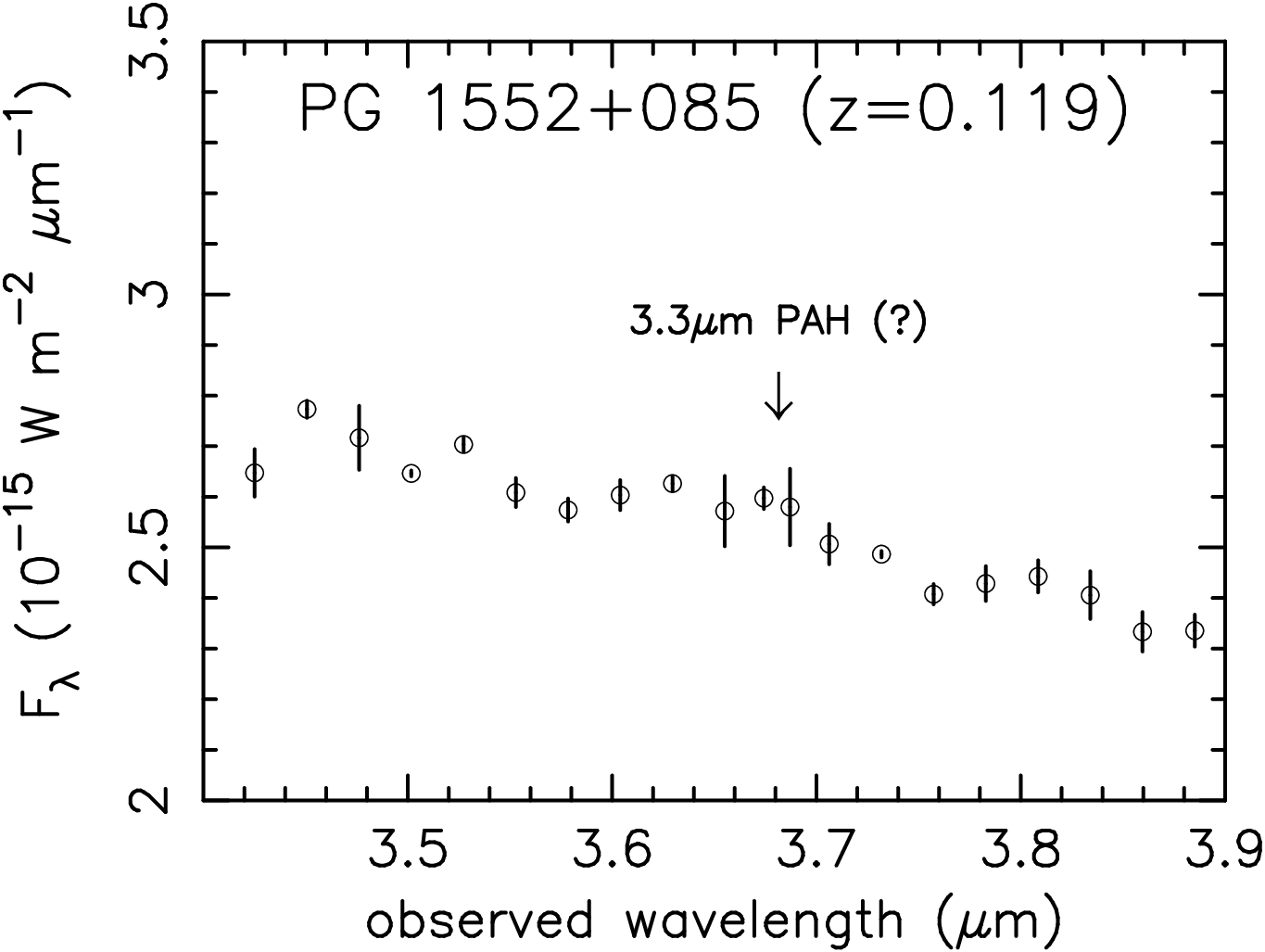} \\
\includegraphics[angle=-90,scale=.185]{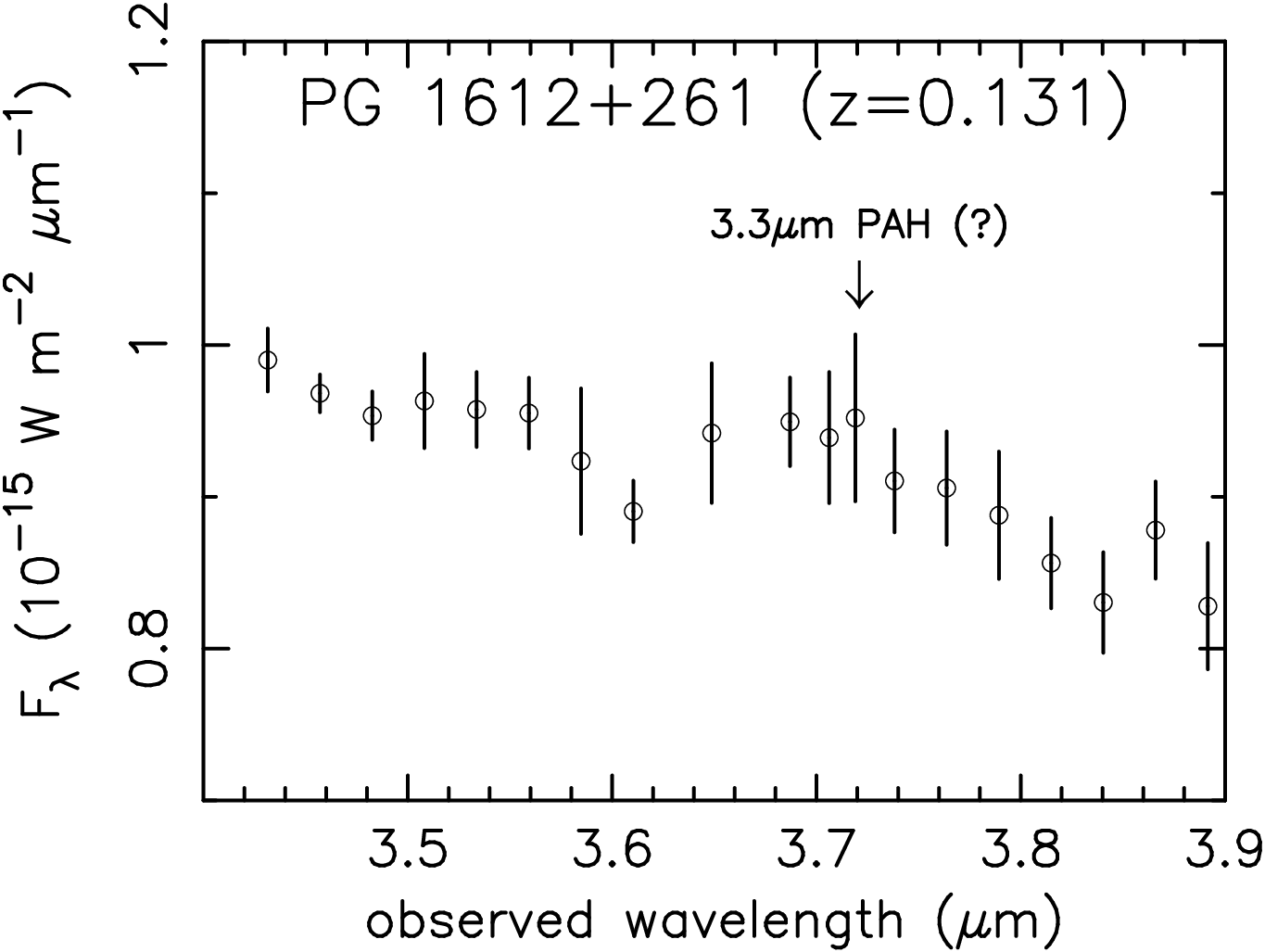} 
\includegraphics[angle=-90,scale=.185]{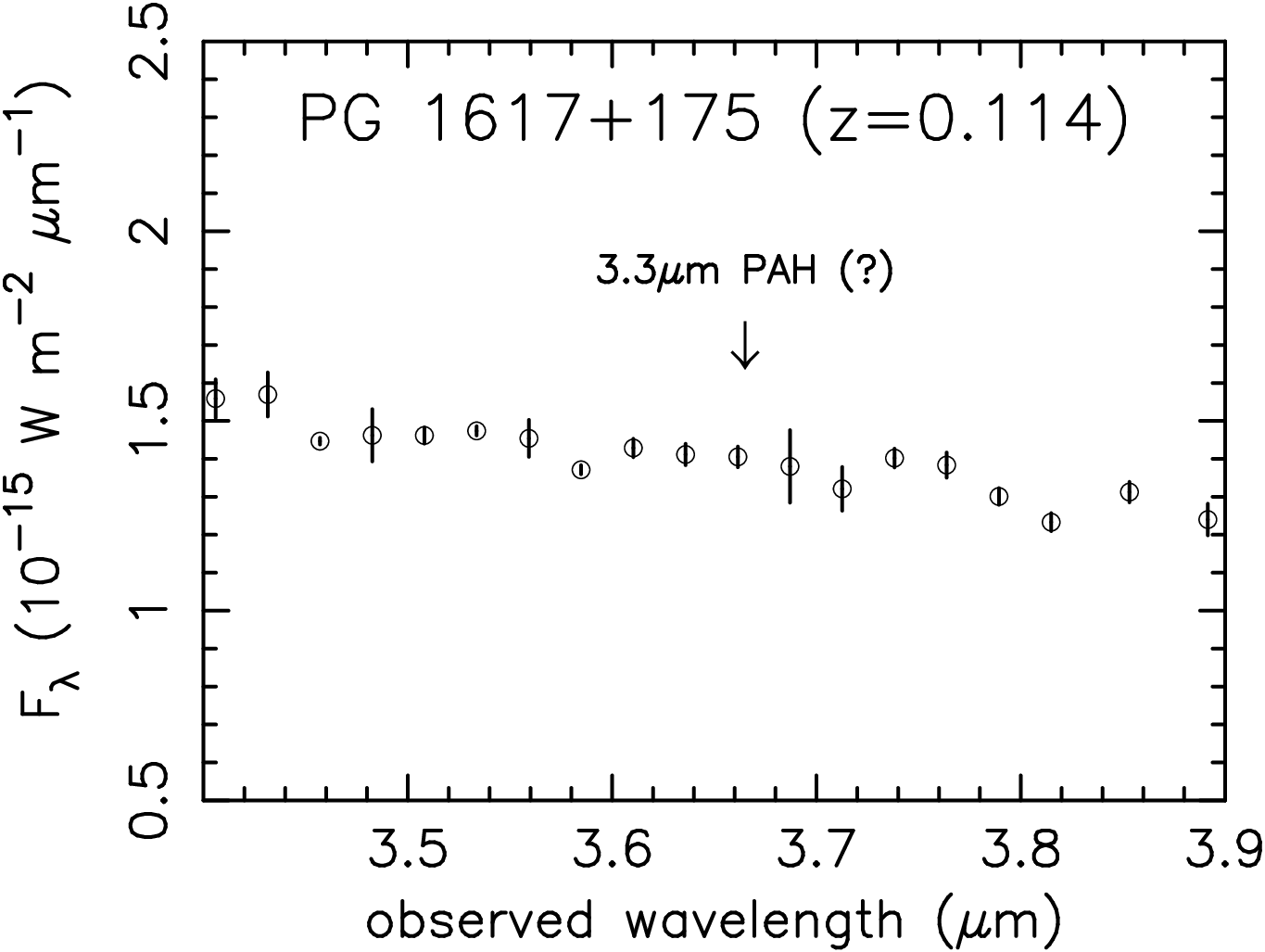} \\
\includegraphics[angle=-90,scale=.185]{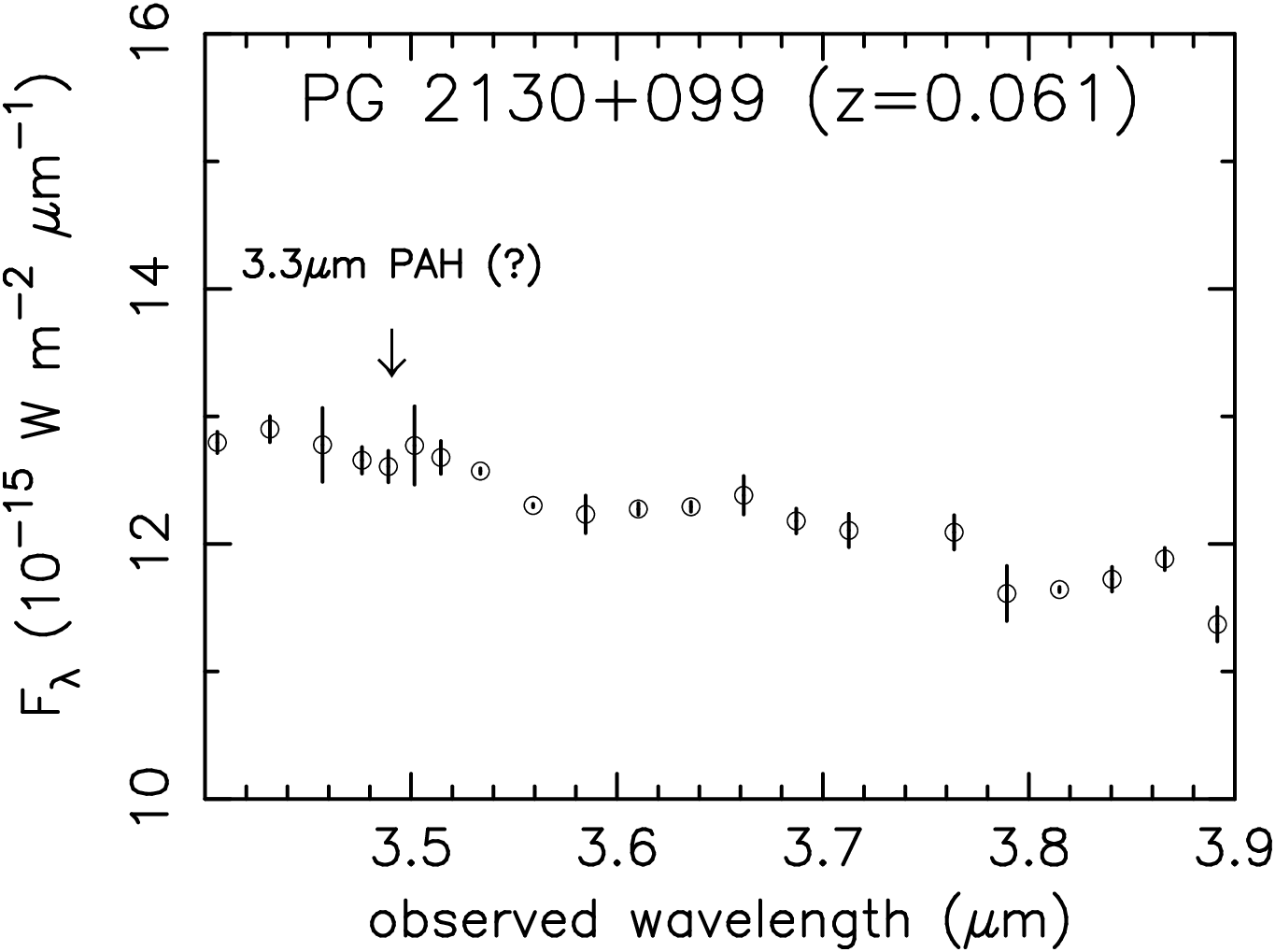} 
\includegraphics[angle=-90,scale=.185]{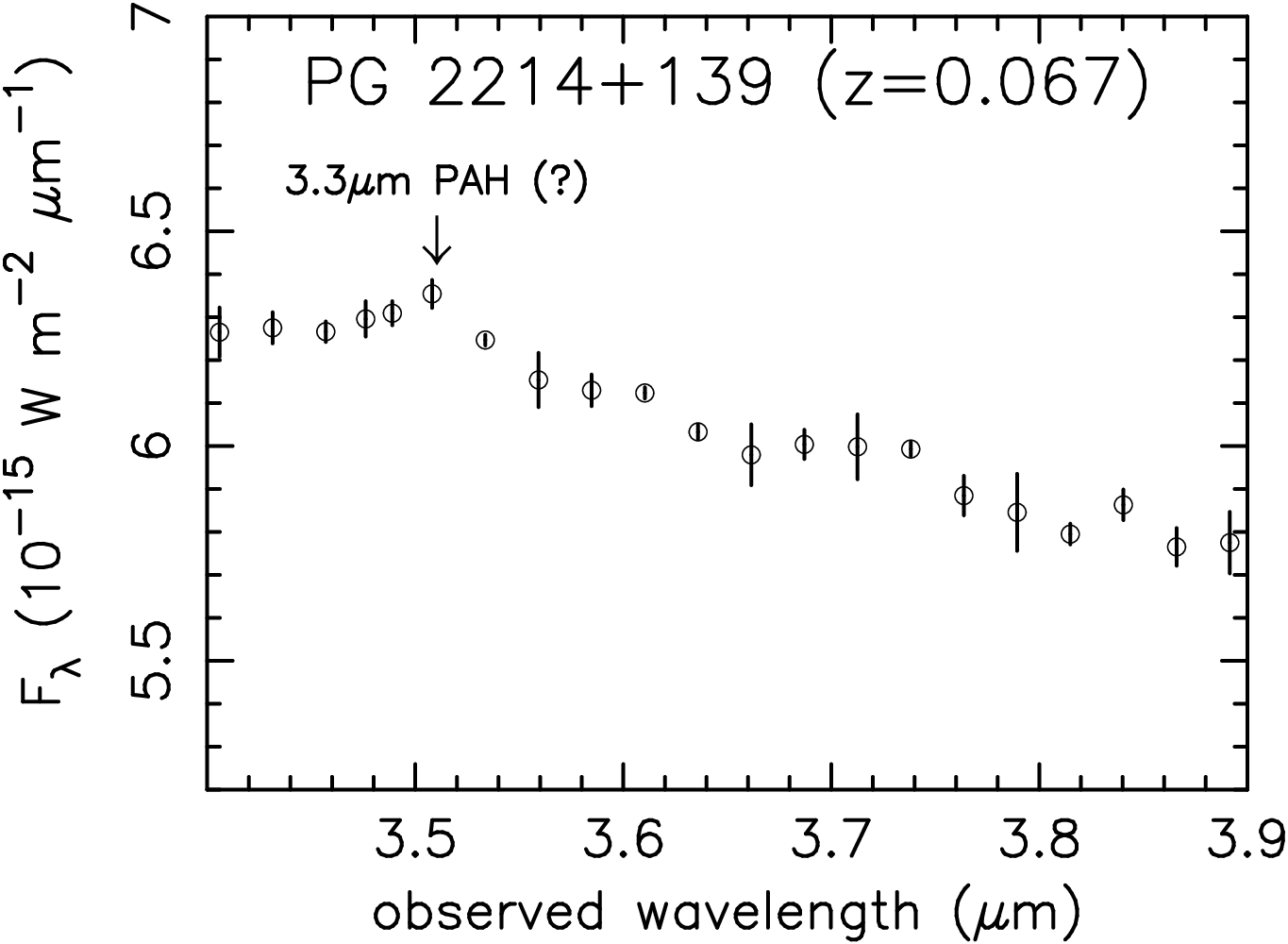} 
\caption{
Infrared 3.4--3.9 $\mu$m spectra of low-luminosity PG QSOs.
Symbols are the same as in Figure 1.
}
\end{figure}

%---  Figure 3 ---%
\begin{figure}
\begin{center}
\includegraphics[angle=-90,scale=.185]{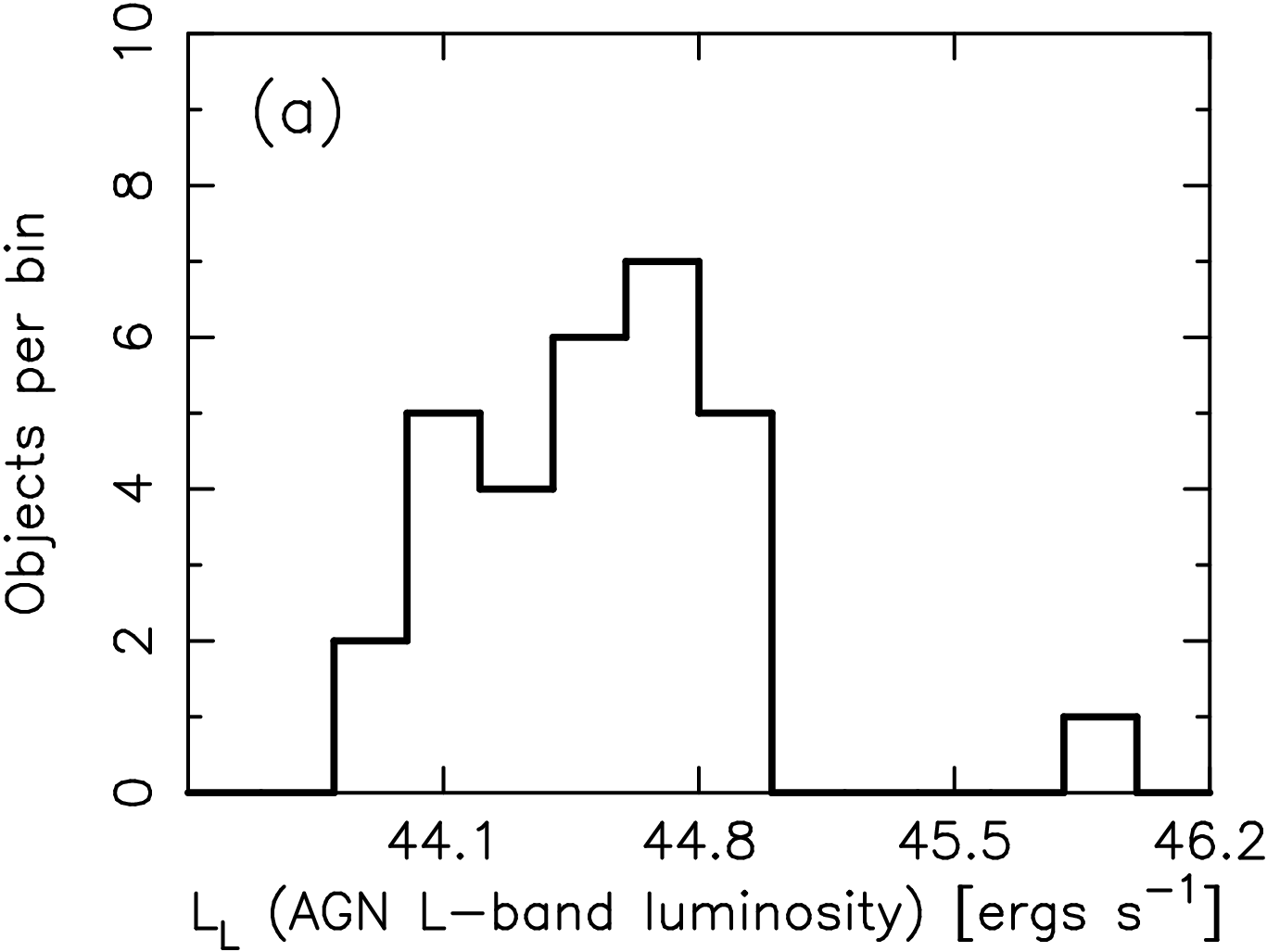} 
\includegraphics[angle=-90,scale=.185]{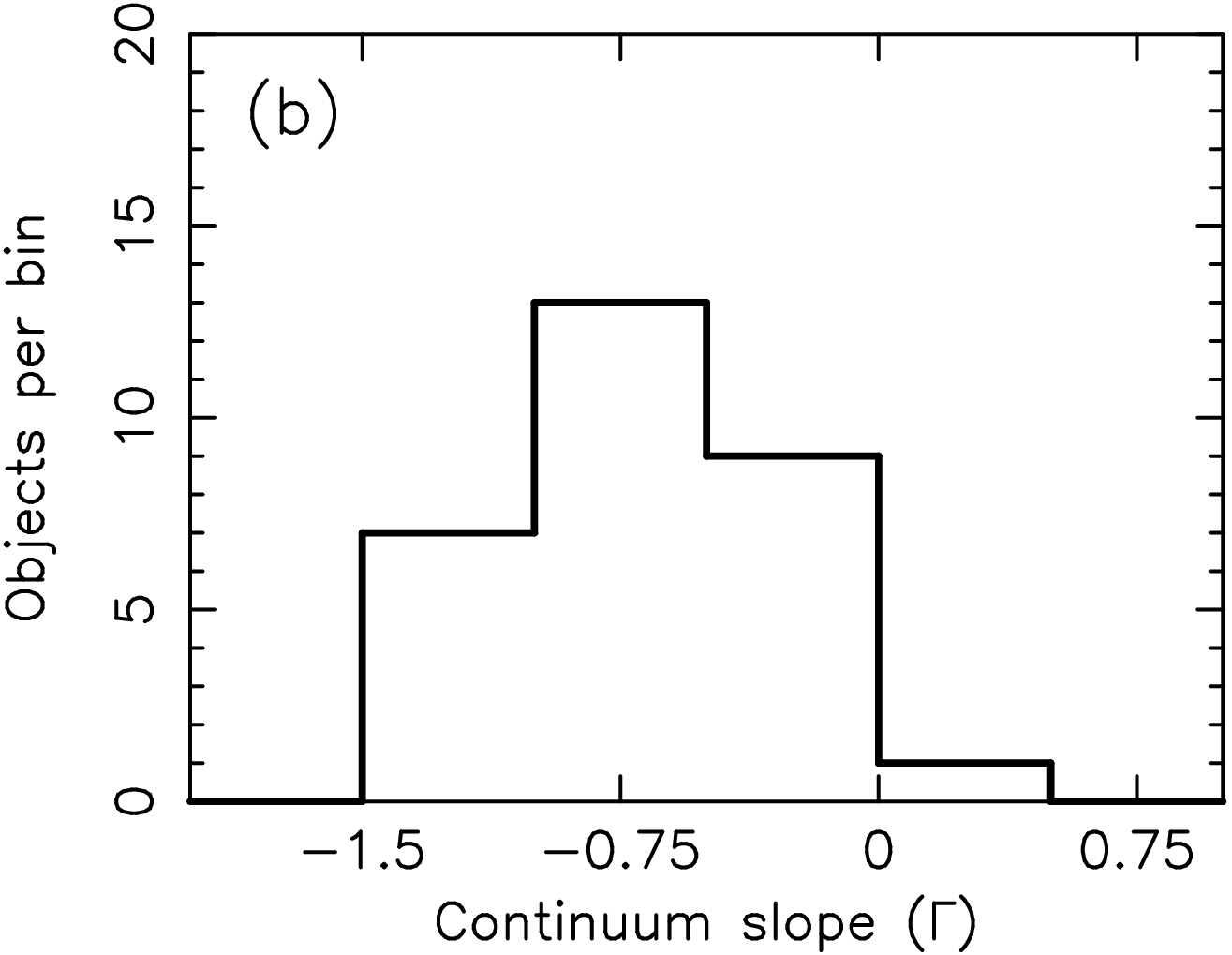} 
\includegraphics[angle=-90,scale=.185]{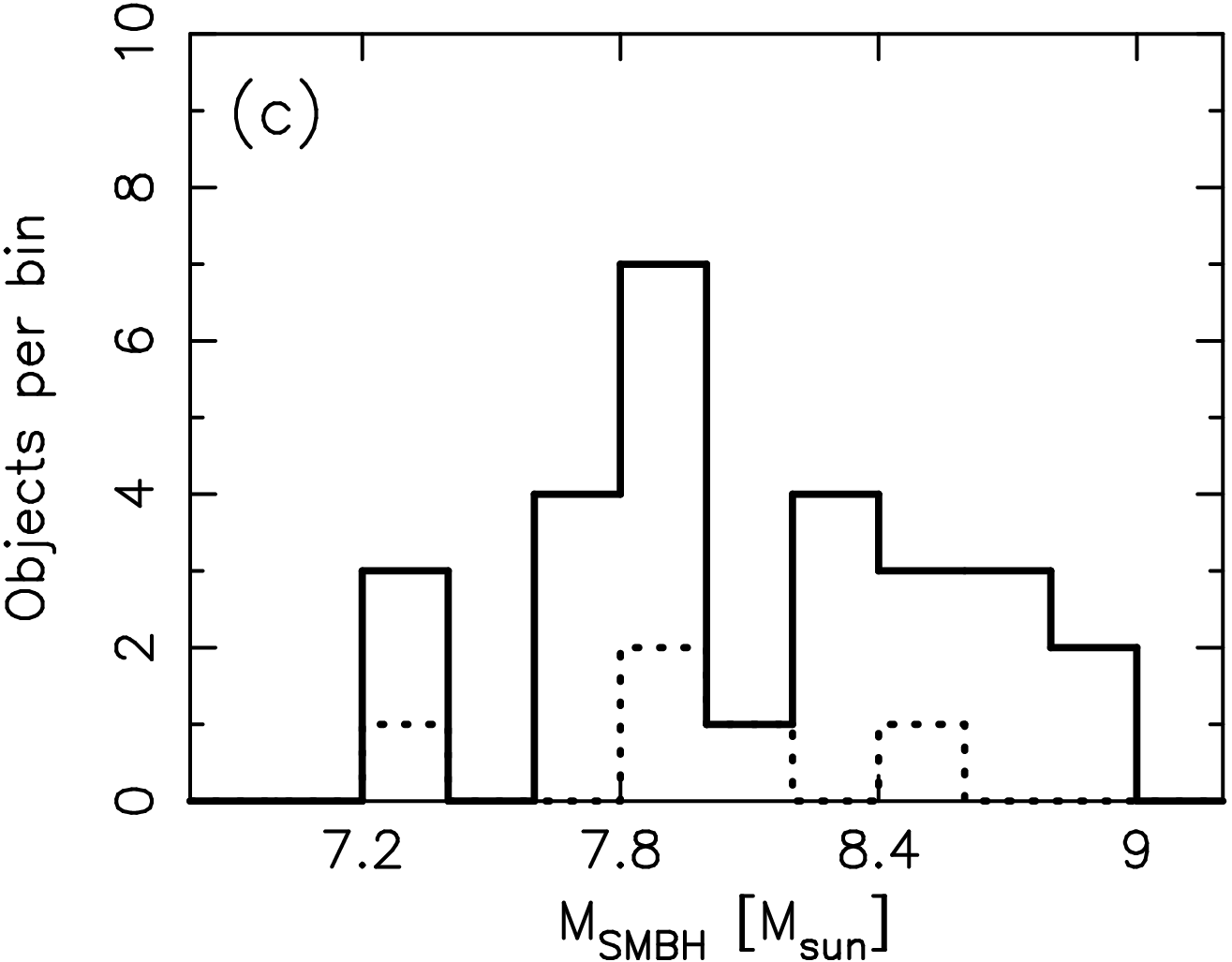} 
\end{center}
\caption{
{\it (a) : } Histogram of the AGN luminosities derived from 
continuum luminosities at $\lambda_{\rm rest}$ = 3.35 $\mu$m 
($\lambda$L$_{\lambda}$) in [ergs s$^{-1}$] for the 30 observed PG
QSOs. 
{\it (b) : } Histogram of the continuum slopes $\Gamma$ (F$_{\lambda}$ 
$\propto$ $\lambda^{\Gamma}$).
{\it (c) : } Histogram of the supermassive black hole masses in units of 
solar mass.
The solid line represents the 27 PG QSOs with available supermassive black hole
masses.
The dotted line represents the five PAH-detected sources. 
}
\end{figure}

%---  Figure 4 ---%
\begin{figure}
\begin{center}
\includegraphics[angle=-90,scale=.35]{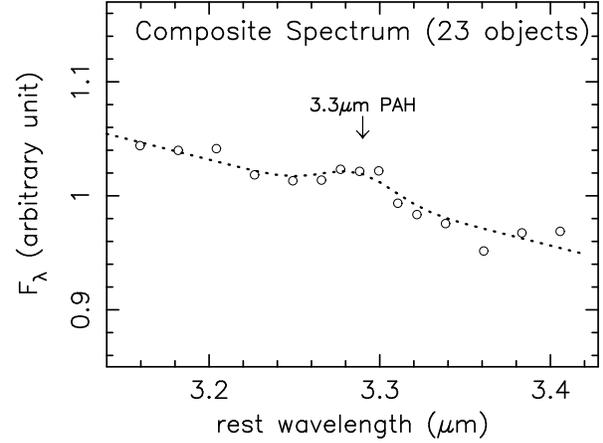} 
\end{center}
\caption{
Composite spectrum of the 23 PAH-undetected PG QSOs observed
with the Subaru IRCS.
The dotted line represents the fitted result, assuming the typical 3.3 $\mu$m
PAH emission profile ($\S$4).
}
\end{figure}

%---  Figure 5 ---%
\begin{figure}
\begin{center}
\includegraphics[angle=-90,scale=.35]{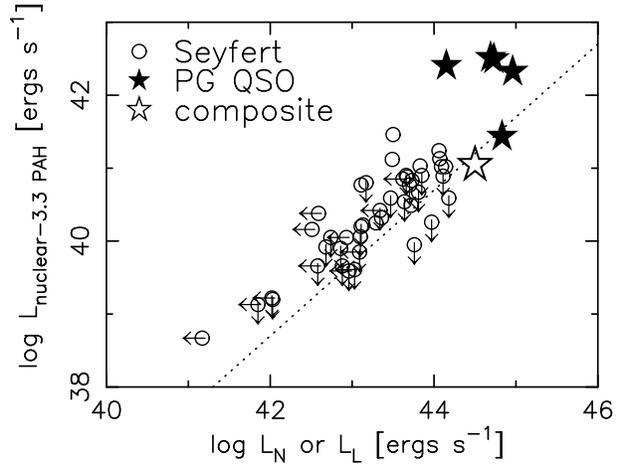} 
\end{center}
\caption{
The ordinate is the nuclear 3.3 $\mu$m PAH emission luminosity in 
[ergs s$^{-1}$].
The abscissa is the nuclear $N$-band (10.8 $\mu$m) luminosity 
($\lambda$L$_{\lambda}$) for Seyferts and the $L$-band (3.35 $\mu$m) luminosity
($\lambda$L$_{\lambda}$) for PG QSOs in [ergs s$^{-1}$].
The ordinate and abscissa trace the nuclear-starburst and AGN-heated dust 
emission luminosity, respectively.
Open circles represent Seyfert galaxies studied by \citet{iw04}.
New results by \citet{oi10} are added when nuclear $N$-band photometric
measurements are available.
For the Seyfert galaxies observed in both papers, new results by 
\citet{oi10} are adopted.
Filled stars represent PG QSOs with detectable 3.3 $\mu$m PAH emission in
their individual spectra. 
The open star represents the composite spectrum of the 23 PAH-undetected PG
QSOs, for which we adopt the median $\lambda$L$_{\lambda}$(3.35 $\mu$m)
value of 3 $\times$ 10$^{44}$ ergs s$^{-1}$ in the abscissa. 
The dotted line represents a nuclear-starburst-to-AGN bolometric luminosity
ratio of 0.1 (see $\S$5). 
}
\end{figure}


\clearpage

\begin{thebibliography}{}
% Journals(e.g. A\&A,ApJ,AJ,NMRAS,PASP ...)
% Authors, Year, Journal, Vol#, Page#
% Journal Title Abbreviation >> http://www.asj.or.jp/pasj/Jabb.html
\bibitem[Antonucci \& Millar(1985)]{ant85}
         Antonucci, R. R. J. \& Millar, J. S. 1985, ApJ, 297, 621 
\bibitem[Ballantyne(2008)]{bal08}
         Ballantyne, D. R. 2008, ApJ, 685, 787
\bibitem[Barvainis(1987)]{bar87}
         Barvainis, R. 1987, ApJ, 320, 537
\bibitem[Baum et al.(2010)]{bau10}
         Baum, S. T., et al. 2010, ApJ, 710, 289
\bibitem[Collin \& Zahn(2008)]{col08}
         Collin, S., \& Zahn, J. -P. 2008, A\&A, 477, 419  
\bibitem[Ferrarese \& Merritt(2000)]{fer00}
         Ferrarese, L., \& Merritt, D. 2000, ApJ, 539, L9 
\bibitem[Genzel et al.(1998)]{gen98} 
         Genzel, R. et al. 1998, ApJ, 498, 579 
\bibitem[Glikman et al.(2006)]{gli06}
         Glikman, E., Helfand, D. J., \& White, R. L. 2006, ApJ, 640,
         579  
\bibitem[Gorjian et al.(2004)]{gor04}
         Gorjian, V. Werner, M. W., Jarrett, T. H., Cole, D. M., \&
         Ressler, M. E. 2004, ApJ, 605, 156
\bibitem[Houck et al.(2004)]{hou04} 
         Houck, J. R., et al. 2004, ApJS, 154, 18
\bibitem[Huchra \& Burg(1992)]{huc92}
         Huchra, J., \& Burg, R. 1992, ApJ, 393, 90
\bibitem[Imanishi(2002)]{ima02} 
         Imanishi, M. 2002, ApJ, 569, 44
\bibitem[Imanishi(2003)]{ima03}
         Imanishi, M. 2003, ApJ, 599, 918
\bibitem[Imanishi \& Dudley(2000)]{imd00} 
         Imanishi, M., \& Dudley, C. C. 2000, ApJ, 545, 701 
\bibitem[Imanishi et al.(2006)]{idm06}
         Imanishi, M., Dudley, C. C., \& Maloney, P. R. 2006, ApJ, 637, 
         114
\bibitem[Imanishi et al.(2010)]{ima10}
         Imanishi, M., Nakagawa, T., Shirahata, M., Ohyama, Y., \&
         Onaka, T. 2010, ApJ, 721, 1233
\bibitem[Imanishi \& Wada(2004)]{iw04} 
         Imanishi, M., \& Wada, K. 2004, ApJ, 617, 214
\bibitem[Iye et al.(2004)]{iye04} 
         Iye, M. et al., 2004, PASJ, 56, 381
\bibitem[Kawakatu et al.(2007)]{kaw07}
         Kawakatu, N., Imanishi, M., \& Nagao, T. 2007, ApJ, 661, 660
\bibitem[Kawakatu et al.(2003)]{kaw03}
         Kawakatu, N., Umemura, M., \& Mori, M. 2003, ApJ, 583, 85
\bibitem[Kawakatu \& Wada(2008)]{kaw08}
         Kawakatu, N., \& Wada, K. 2008, ApJ, 681, 73
\bibitem[Kobayashi et al.(2000)]{kob00}
         Kobayashi, N., et al. 2000, IRCS: Infrared camera and spectrograph 
         for the Subaru Telescope, in Proc. SPIE 4008: Optical and IR
         Telescope Instrumentation and Detectors, eds M. Iye \&
         A. F. Moorwood, 1056  
\bibitem[Magorrian et al.(1998)]{mag98}
         Magorrian, J., et al. 1998, ApJ, 115, 2285 
\bibitem[Moorwood(1986)]{moo86}
         Moorwood, A. F. M. 1986, A\&A, 166, 4
\bibitem[Mouri et al.(1990)]{mou90}
         Mouri, H., Kawara, K., Taniguchi, Y., \& Nishida, M.
         1990, ApJ, 356, L39
\bibitem[Neugebauer et al.(1987)]{neu87}
         Neugebauer, G., Green, R. F., Matthews, K., Schmidt, M.,
         Soifer, B. T., \& Bennett, J. 1987, ApJS, 63, 615
\bibitem[Nishiyama et al.(2008)]{nis08}
         Nishiyama, S., Nagata, T., Tamura, M., Kandori, R., Hatano, H.,
         Sato, S., \& Sugitani, K. 2008, ApJ, 680, 1174
\bibitem[Nishiyama et al.(2009)]{nis09}
         Nishiyama, S., Tamura, M., Hatano, H., Kato, D., Tanabe, T.,
         Sugitani, K., \& Nagata, T. 2009, ApJ, 696, 1407
\bibitem[Norman \& Scoville(1988)]{nor88}
         Norman, C., \& Scoville, N. 1988, ApJ, 332, 124
\bibitem[Ohsuga et al.(1999)]{ohs99}
         Ohsuga, K., Umemura, M., Fukue, J., \& Mineshige, S. 1999,
         PASJ, 51, 345
\bibitem[Oi et al.(2010)]{oi10}
         Oi, N., Imanishi, M., \& Imase, K. 2010, PASJ, 62, 1509
\bibitem[Peeters et al.(2004)]{pee04}
         Peeters, E., Spoon, H. W. W., \& Tielens, A. G. G. M. 2004,
         ApJ, 613, 986
\bibitem[Rayner et al.(2003)]{ray03}
         Rayner, J. T., Toomey, D. W., Onaka, P. M., Denault, A. J.,
         Stahlberger, W. E., Vacca, W. D., Cushing, M. C., \& Wang, S.
         2003, PASP, 115, 362
\bibitem[Risaliti et al.(2010)]{ris10}
         Risaliti, G., Imanishi, M., \& Sani, E. 2010, MNRAS, 401, 197 
\bibitem[Roche et al.(1991)]{roc91}
         Roche, P. F., Aitken, D. K., Smith, C. H., \& Ward, M. J., 
         1991, MNRAS, 248, 606
\bibitem[Rodriguez-Ardila \& Viegas(2003)]{rod03}
         Rodriguez-Ardila, A., \& Viegas, S. M. 2003, MNRAS, 340, L33
\bibitem[Sanders et al.(1989)]{san89}
         Sanders, D. B., Phinney, E. S., Neugebauer, G., Soifer, B. T.,
         Matthews, K. 1989, ApJ, 347, 29
\bibitem[Schmidt \& Green(1983)]{sch83}
         Schmidt, M., \& Green, R. F. 1983, ApJ, 269, 352
\bibitem[Schweitzer et al.(2006)]{sch06}
         Schweitzer, M., et al. 2006, ApJ, 649, 79
\bibitem[Shi et al.(2007)]{shi07}
         Shi, Y., et al. 2007, ApJ, 669, 841
\bibitem[Shure et al.(1994)]{shu94} 
         Shure, M. A., Toomey, D. W., Rayner, J. T., Onaka, P., \& 
         Denault, A. J. 1994, Proc. SPIE, 2198, 614
\bibitem[Smith et al.(2007)]{smi07} 
         Smith, J. D., et al. 2007, ApJ, 656, 770
\bibitem[Soifer et al.(2002)]{soi02}
         Soifer, B. T., Neugebauer, G., Matthews, K., Egami, E., 
         \& Weinberger, A. J. 2002, AJ, 124, 2980 
\bibitem[Spinoglio \& Malkan(1989)]{spi89}
         Spinoglio, L., \& Malkan, M. A. 1989, ApJ, 342, 83
\bibitem[Tielens(2008)]{tie08}
         Tielens, A. G. G. M. 2008, ARA\&A, 46, 289 
\bibitem[Tokunaga(2000)]{tok00}
         Tokunaga, A. T. 2000, in Allen's
         Astrophysical Quantities, ed.
         A. N. Cox (4th ed; Berlin: Springer), 143
\bibitem[Tokunaga et al.(1991)]{tok91}
         Tokunaga A. T., Sellgren K., Smith R. G., Nagata T., 
         Sakata A., Nakada Y., 1991, ApJ, 380, 452 
\bibitem[Umemura et al.(1997)]{ume97}
         Umemura, M., Fukue, J., \& Mineshige, S. 1997, ApJ, 479, L97
\bibitem[Umemura et al.(1998)]{ume98}
         Umemura, M., Fukue, J., \& Mineshige, S. 1998, MNRAS, 299, 1123
\bibitem[Veilleux et al.(2009)]{vei09}
         Veilleux, S., et al. 2009, ApJS, 182, 628 
\bibitem[Voit(1992)]{voi92}
         Voit, G. M. 1992, MNRAS, 258, 841
\bibitem[Vollmer \& Beckert(2003)]{vol03}
         Vollmer, B., \& Beckert, T. 2003, A\&A, 404, 21
\bibitem[von Linden et al.(1993)]{von93}
         von Linden, S., Biermann, P. L., Duschl, W, J., Lesch, H., \&
         Schmutzler, T., 1993, A\&A, 280, 468 
\bibitem[Wada \& Norman(2002)]{wad02}
         Wada, K., \& Norman, C. A. 2002, ApJ, 566, L21
\bibitem[Wada et al.(2009)]{wad09}
         Wada, K., Papadopoulos, P. P., \& Spaans, M. 2009, ApJ, 702, 63
\bibitem[Watabe et al.(2008)]{wat08}
         Watabe, Y., Kawakatu, N., \& Imanishi, M. 2008, ApJ, 677, 895
\bibitem[Watabe et al.(2009)]{wat09}
         Watabe, Y., Kawakatu, N., Imanishi, M., \& Takeuchi, T.T. 2009,
         MNRAS, 400, 1803 
\bibitem[Wu et al.(2007)]{wu07}
         Wu, X-B. 2007, ApJ, 657, 177
\bibitem[Wu et al.(2009)]{wu09}
         Wu, Y., Charmandaris, V., Huang, J., Spinoglio, L., \&
         Tommasin, S. 2009, ApJ, 701, 658
\end{thebibliography}
\end{document}